\documentclass[a4paper,11pt]{article}

\usepackage{graphicx,array}
\usepackage{color}
\usepackage{latexsym}
\usepackage{amsthm}
\usepackage{amsmath}
\usepackage{amssymb}
\usepackage{empheq}

\usepackage{dsfont}
\usepackage{epsfig}
\usepackage{slashed}
\usepackage{bbold}
\usepackage{psfrag}
\usepackage[svgnames]{xcolor}
\PassOptionsToPackage{caption=false}{subfig}
\usepackage{subcaption}
\usepackage{xfrac}
\usepackage{multirow}
\usepackage{booktabs}
\usepackage{cite}
\usepackage[normalem]{ulem}
\usepackage{jheppub} 
\usepackage{soul}

\newcommand{\be}{\begin{equation}}
\newcommand{\ee}{\end{equation}}
\newcommand{\bea}{\begin{eqnarray}}
\newcommand{\eea}{\end{eqnarray}}
\newcommand{\bei}{\begin{itemize}}
\newcommand{\eei}{\end{itemize}}

\title{Probing right-handed neutrinos dipole operators}

\author[a,b]{Daniele Barducci}
\author[c]{Enrico Bertuzzo}
\author[d]{Marco Taoso}
\author[a,b]{Claudio Toni}
\affiliation[a]{Universit\`a degli Studi di Roma la Sapienza, Piazzale Aldo Moro 5, 00185, Roma, Italy}
\affiliation[b]{INFN Section of Roma 1, Piazzale Aldo Moro 5, 00185, Roma, Italy}
\affiliation[c]{Instituto de Fisica, Universidade de Sao Paulo, C.P. 66.318, 05315-970 Sao Paulo, Brazil}
\affiliation[d]{Istituto Nazionale di Fisica Nucleare, Sezione di Torino, via P. Giuria 1, I–10125
Torino, Italy}
\emailAdd{daniele.barducci@roma1.infn.it}
\emailAdd{bertuzzo@if.usp.br}
\emailAdd{claudio.toni@uniroma1.it}
\emailAdd{marco.taoso@to.infn.it}

\abstract{We consider the minimal see-saw extension of the Standard Model with two right-handed singlet fermions $N_{1,2}$ with mass at the GeV scale, augmented by an effective dipole operator between the sterile states. We firstly review current bounds on this effective interaction from fixed-target and collider experiments, as well as from astrophysical and cosmological observations. We then highlight the prospects for testing the decay $N_2 \to N_1 \gamma$ induced by the dipole at future facilities targeting long-lived particles such as ANUBIS, CODEX-b, FACET, FASER 2, MAPP and SHiP.}

\begin{document} 
\maketitle

\section{Introduction and framework}\label{sec:intro}

The see-saw mechanism~\cite{Minkowski:1977sc,Mohapatra:1979ia,Yanagida:1979as,Gell-Mann:1979vob,Schechter:1981bd} is arguably the simplest extension of the Standard Model (SM) that is able to explain the observed pattern of neutrino masses and oscillations. In its simplest incarnation, it consists in adding to the SM particle content a right-handed (RH) neutrino, that is a spin 1/2 fermion, singlet under the SM gauge group, which has a Yukawa interaction with SM leptons, as well as a Majorana mass term. One of the active neutrinos acquire thus a non-vanishing mass $m_\nu$ and a mixing $\theta$ with the new sterile state, parametrically expressed by the relations
\be\label{eq:naive_seesaw}
m_\nu \simeq \frac{y_\nu^2 v^2}{m_N} \ , \quad\quad \theta \simeq \sqrt{\frac{m_\nu}{m_N}}, 
\ee
where $y_\nu$ and $m_N$ are the RH neutrino Yukawa interaction and mass respectively and $v$ is the electroweak (EW) vacuum expectation value (VEV). Since experimental data point to at least two massive neutrinos, at least two RH states must be added to obtain a realistic phenomenology. In this case the essence of the see-saw mechanism is unaltered, with the obvious promotion of $y_\nu$ and $m_N$ to matrices in flavor space, but the relations of Eq.~\eqref{eq:naive_seesaw} turn out to be modified. In particular the mixing angles can receive an {\emph{exponential enhancement}} with respect to the naive see-saw scaling case, that may drastically modify the phenomenology. This is best seen in the Casas-Ibarra parametrization \cite{Casas:2001sr}. From the practical point of view, this means that masses and mixings can be treated as independent parameters.
Irrespective of this consideration, by fixing $m_\nu$ Eq.~\eqref{eq:naive_seesaw} doesn't uniquely point to a preferred mass range for $m_N$, which could lie all the way up the grand unification scale if $y_\nu$ is an ${\cal O}(1)$ parameter. However, in recent years RH neutrinos with mass below the EW scale have gained more and more attention in that they can explain the matter-antimatter asymmetry of the Universe via neutrino oscillations~\cite{Akhmedov:1998qx,Asaka:2005pn} and, crucially, can be tested at  present and future colliders and fixed-target experiments, see e.g.~\cite{Keung:1983uu,Ferrari:2000sp,delAguila:2008cj,BhupalDev:2012zg,Helo:2013esa,Blondel:2014bra,Abada:2014cca,Cui:2014twa,Antusch:2015mia,Gago:2015vma,Antusch:2016vyf,DeRomeri:2016gum,Caputo:2016ojx,Gago:2016wzr,Abada:2017jjx,Antusch:2017hhu,Das:2017zjc,Abada:2018sfh,Cottin:2018nms,Helo:2018qej,Cottin:2018kmq,Hernandez:2018cgc,Dercks:2018wum,Jones-Perez:2019plk,Hirsch:2020klk,Zhou:2021ylt,DeVries:2020jbs}. 

While the see-saw model is a full-fledged ultraviolet (UV) complete theory, at least in the same way as the SM is, in the case where RH neutrinos lie at the EW scale it is interesting to consider it as a low energy effective field theory (EFT) extended with higher dimensional operators built from the SM and the RH neutrino fields. The resulting theory is called $\nu$SMEFT and is described by the following Lagrangian
\be\label{eq:lag_nusmeft}
{\cal L} = {\cal L}_{\rm SM}  + i \bar N \slashed \partial N - \bar L_L Y_\nu \tilde H N - \frac{1}{2}  \bar N^c M_{N} N + \sum_{n>4} \frac{{\cal O}^n}{\Lambda^{n-4}} + h.c. \ ,
\ee
where $N$ is a vector describing ${\cal N}_f$ flavors of RH neutrino fields and $N^c = C \bar N^T$, with $C= i \gamma^2 \gamma^0$. Furthermore, $Y_\nu$ is the $3\times {\cal N}_f$ Yukawa matrix of the neutrino sector with $\tilde H = i \sigma^2 H^*$, $M_{N}$ is a ${\cal N}_f\times {\cal N}_f$ Majorana mass matrix for the RH neutrinos  and ${\cal O}^n$ the Lorentz and gauge invariant operators with dimension $n$ built out from the SM and the RH neutrino fields, with $\Lambda$ parametrizing the Wilson coefficient of the operator.
A complete and independent set of operators has been built up to dimension nine~\cite{delAguila:2008ir,Liao:2016qyd,Li:2021tsq}. Interestingly, already at $d=5$ two genuine $\nu$SMEFT operators appear\footnote{The other $d=5$ operator is clearly the Weinberg operator ${\cal O}^5_W=(\bar L^c \tilde H^*)(\tilde H^\dag L)$.}. 
The first is an operator coupling the RH neutrinos with the Higgs boson, ${\cal O}^5_{NH}=\bar N^c N H^\dag H$. This triggers a new decay mode for the Higgs, with interesting consequences for collider phenomenology, both at the Large Hadron Collider (LHC)~\cite{Graesser:2007yj,Graesser:2007pc,Caputo:2017pit,Butterworth:2019iff} and future colliders~\cite{Barducci:2020icf}. The second operator is a dipole with the 
 hypercharge gauge boson\footnote{We define $\sigma^{\mu\nu}=i[\gamma_\mu,\gamma_\nu]/2$.} ${\cal O}^5_{NB}=\bar N^c \sigma^{\mu\nu} N B_{\mu\nu}$, which has so far been less investigated~\cite{Aparici:2009fh,Balaji:2020oig,Barducci:2020ncz,Barducci:2020icf,Cho:2021yxk,Delgado:2022fea}\footnote{
Recent works on the phenomenology of d=6 operators involving sterile neutrinos at accelerators are~\cite{Alcaide:2019pnf,Han:2020pff,Zhou:2021ylt,DeVries:2020jbs,Beltran:2021hpq,Cottin:2021lzz}}. 
Among other effects, this operator generates the decay\footnote{Given the mass range that we consider, the decay process in which the $\gamma$ is substituted with a $Z$ boson is kinematically closed.} 
\be\label{eq:decay}
N_{\rm heavy} \to N_{\rm light} + \gamma \ , \quad m_{N_{\rm heavy}}>m_{N_{\rm light}}.
\ee
 This interaction is the subject of our study.
 Our focus will be on light RH neutrinos with masses up to a few GeV. Such light states can be produced not only at high energy colliders via parton interactions, but also at fixed-target experiments, typically via meson decay.
More specifically, we will analyze in detail the current bounds from colliders experiments, such as LHC, LEP and BaBar, and fixed-target experiments, such as CHARM~\cite{CHARM:1985anb}, NuCal~\cite{Blumlein:1990ay,Blumlein:1991xh} and NA64~\cite{Bernhard:2020vca}. We will then compute the predicted sensitivity to the $\nu$SMEFT parameter space of the proposed experiments ANUBIS~\cite{Bauer:2019vqk} , CODEX-b~\cite{Gligorov:2017nwh,Aielli:2019ivi,Aielli:2022awh}, FACET~\cite{Cerci:2021nlb}, FASER 2~\cite{Feng:2017uoz,Feng:2022inv}, MAPP~\cite{Staelens:2019gzt,Pinfold:2019zwp} and SHiP~\cite{SHiP:2015vad,SHiP:2021nfo}. In addition, we will also discuss constraints from cosmology and astrophysics.

Throughout this work we will consider the theory of Eq.~\eqref{eq:lag_nusmeft}, focusing on the $d=5$ dipole operator.
Given its symmetry properties, this operator is non-vanishing only for ${\cal N}_f\ge 2$. Since we are primarily interested in probing the effect of the dipole operator, we will work under the assumption that the active-sterile mixing effects are negligible for what concerns the {\emph{heavier}} sterile neutrinos phenomenology, in such a way that their decay proceeds only via the dipole operator under our scrutiny through the process of Eq.~\eqref{eq:decay}.
As for the lightest RH neutrino $N_1$, its decay pattern is completely determined by the active-sterile mixing, as in the standard see-saw case. As we are going to discuss in Sec. \ref{sec:currentbounds}, the $N_1$ lifetime can be strongly constrained by cosmological observations, especially the ones related to the epoch of Big Bang Nucleosynthesis (BBN). A relatively safe scenario is the one in which $N_1$ mixes dominantly with the third generation of SM neutrinos, $\nu_\tau$. We found that this configuration can be easily obtained by choosing ${\cal N}_f =3$, satisfying at the same time all other relevant constraints. Interestingly, in this case the heaviest RH neutrinos $N_3$ can be decoupled from the spectrum without affecting the mixing pattern for $N_1$, leaving only the two lightest RH neutrinos $N_{1,2}$ as dynamical states. In presenting our main findings we will thus consider a framework with only these two states living at the EW scale and interacting via the dipole operator which we normalize as
\be\label{eq:dipole_op}
{\cal O}^5_{NB} = \frac{g_Y}{16\pi^2} \frac{e^{i \alpha}}{\Lambda} \bar N_{1}^c \sigma^{\mu\nu} N_{2} B_{\mu \nu} + h.c. \ ,
\ee
where $m_{N_2}>m_{N_1}$, and where $g_Y$ and $B_{\mu\nu}$ are the $U(1)_Y$ coupling and field strength tensor respectively. The loop suppression factor is explicitly introduced since this operator only arises at loop level in any weakly coupled UV completion, see {\emph{e.g.}}~\cite{Buchmuller:1985jz,Craig:2019wmo}, while the hypercharge coupling is added because of the presence of $B_{\mu\nu}$. Explicit UV completions include models with additional scalar and fermions or models with additional vectors and fermions, with non-vanishing hypercharge \cite{Aparici:2009fh,Aparici:2009oua}. We will comment later on possible strongly interacting UV completions. Since the Wilson coefficient can be complex, we show explicitly its phase $\alpha$. In this scenario, ${\cal O}^5_{NB}$ completely governs the RH neutrinos phenomenology\footnote{Our analysis focuses on the radiative decays of $N_2$ induced by the dipole operator but, of course, additional signals at the experiments under study could be produced by $N_1$ decays, if the mixing with the active sector is not too suppressed.}. In particular, it dictates the heaviest neutrino $N_2$ total decay width. For RH neutrinos below the $Z$ mass the dominant decay mode is $N_2\to N_1\gamma$ whose rate reads
\be\label{eq:N2decay}
\Gamma(N_2 \to N_1\gamma) = \frac{{g_Y}^2}{(16\pi^2)^2} \frac{c_w^2}{2\pi}  \frac{m_{N_1}^3}{\Lambda^2} \delta^3 \left(\frac{2+\delta}{1+\delta}\right)^3 \simeq \frac{g_Y^2}{64\pi^5} c_w^2\frac{m_{N_1}^3}{\Lambda^2}\delta^3 \ ,
\ee
where $c_w$ is the cosine of the Weinberg angle and the last equality holds for small values of $\delta$, which is defined as
\be
\delta = \frac{m_{N_2}-m_{N_1}}{m_{N_1}} \ .
\ee
The three-body decay into an off-shell $Z$ boson provides a subdominant contribution. 
As it is clear from Eq.~\eqref{eq:N2decay}, the relative mass splitting $\delta$ is crucial in determining the RH neutrino decay length, and hence its lifetime. This gives an indication on
the type of experiments that can have a sensitivity to this scenario, depending on how far the detector is located with respect to the $N_2$ production points. For example the neutrinos $N_2$ could decay promptly, {\emph{i.e.}} with a typical decay length smaller than ${\cal O}({\rm mm})$. In this case they are a primary target for standard collider searches. Their lifetime could also be longer, with corresponding decay lengths in the ${\cal O}(1\;{\rm m}- 100\;{\rm m})$, for which different strategies need to be envisaged. In the more extreme case, they can be stable with respect to the length scale of any terrestrial experiment and hence completely invisible for what concerns laboratory searches. We will comment upon all these possibilities in the following, mainly focusing however on a region of parameter space in which the heavier neutrino $N_2$ is a long-lived state with a macroscopic decay length. This choice has a twofold motivation. From one side, light new states with suppressed interactions, as the one inherited from the dipole operator, have usually a long lifetime. From the other side, the study of long-lived particles is an active field which has received a lot of attention in the last years, following the philosophy of {\emph{leaving no stones unturned}} and {\emph{lighting new lampposts}} in the quest of new physics beyond the SM. In this area, big experimental progresses are foreseen in the mid- and short-term.

The relative mass splitting is also important in determining the photon energy arising from the decay of Eq.~\eqref{eq:decay}. From basic kinematics in the $N_2$ rest frame one has
\be
E_\gamma^{\rm com} = m_{N_2}\frac{\delta}{2} \frac{2+\delta}{(1+\delta)^2} \ .
\ee
Assuming the photon to be produced collinearly with the direction of $N_2$ in the laboratory frame, which maximizes the photon energy  in this frame of reference, one has
\be\label{eq:Egamma_LAB}
E_\gamma^{\rm lab} = \left( P_{N_2} + \sqrt{m_{N_2}^2+P_{N_2}^2}\,\right) \frac{\delta}{2} \frac{2+\delta}{(1+\delta)^2} \simeq 2 P_{N_2} \delta \ ,
\ee
where $P_{N_2}$ is the modulus of the $N_2$ spatial momentum and the last equality holds for $m_{N_2}/P_{N_2}\ll 1$ and $\delta \ll 1$\footnote{Notice that, for vanishing active-sterile mixing, in the $\delta \to 0$ limit the mass term in Eq. \eqref{eq:lag_nusmeft} becomes symmetric under a global SO(2) symmetry that acts on the vector $N = (N_1, N_2)^T$. It is thus technically natural to have small $\delta$.}. Thus the smaller the relative mass splitting the softer the final state photons, which however should satisfy some minimal threshold requirement in order to be identified in a detector. Hence too small relative mass splittings will hardly be experimentally testable.

The rest of the paper is organized as follows. In Sec.~\ref{sec:currentbounds} we review the existing limits on the dipole operator from cosmology, colliders and other type of experiments, while in Sec.~\ref{sec:SHiPFaser} we discuss the future sensitivity of SHiP and FASER 2 on the model parameter space, wrapping up our conclusion in Sec.~\ref{sec:conc}. We also add three appendices. In App.~\ref{App:futureLHC} we present results for other future LHC experiments targeting long-lived particles, in App.~\ref{App:widths} we report useful formul\ae~for computing the decay of a QCD meson into a pair of RH neutrinos via the dipole operator and in App.~\ref{sec:e-recoil} we discuss possible constraints arising from searches of electrons recoil signatures in laboratory experiments.

\section{Current limits from cosmology, colliders and other experiments }
\label{sec:currentbounds}

The parameter space of the simplified scenario considered in this work is spanned by the lighter neutrino mass $m_{N_1}$, the relative mass splitting $\delta$ with the heavier RH neutrino, and the Wilson coefficient of the dipole operator, parametrized by $\Lambda$ and its phase $\alpha$.
This parameter space is already constrained by laboratory data from colliders and past beam dump experiments, as well as by astrophysical and cosmological measurements. In this section we will review the most important and stringent ones. Particular care must be taken in ensuring the validity of the EFT in the various considered processes. The dipole operator in Eq.~\eqref{eq:dipole_op} induces $N_1 N_2$ production through the exchange of a photon or a $Z$ boson. Following Ref.~\cite{Racco:2015dxa}
and assuming couplings of order one, we identify the EFT cut-off scale with $\Lambda,$ and for the EFT to be valid we require
\be\label{eq:validity_EFT}
\sqrt{\hat{s}} < \Lambda,
\ee
where $\hat{s} = (p_{N_1} + p_{N_2})^2$ is the Lorentz invariant energy that enters the vertex. One important production mechanism for light $N_{1,2}$ is via heavy meson decay, that can be copiously produced in fixed-target experiments. In this case the heavy neutrino production proceeds via an $s$-channel $\gamma$ and we have $\hat{s} = m_M^2$, with $m_M$ the meson mass. For higher masses direct production at collider can be relevant. In this case, $\hat{s}$ is the center of mass energy squared of the parton pair that exchange the photon or the $Z$ boson, {\emph{e.g.}} $e^+e^-$ for LEP or $q\bar{q}$ for the LHC. Analogous considerations apply for other production modes, as for example production via photon bremsstrahlung.

\subsection{Fixed-target experiments}

We start our discussion with fixed-target experiments, for which we have 
considered data collected at CHARM~\cite{CHARM:1985anb}, NuCal~\cite{Blumlein:1990ay,Blumlein:1991xh} and NA64~\cite{Bernhard:2020vca}.
We consider the production of RH neutrinos from the decay $M \to N_1 N_2$, with $M$ a vector meson produced at these experiments.~\footnote{We have checked that the amplitude for the decay $P \to N_1 N_2$ mediated by $Z$ boson exchange and with $P$ a pseudoscalar meson vanishes identically. Moreover, we have estimated that the decay $P \to N_1 N_2\gamma $ provides only a marginal improvement of our sensitivities. For this reason, we do not consider this contribution.}

\paragraph{CHARM:} In the CHARM experiment, a 400\;GeV proton beam was dumped on a copper target. The detector, placed at a distance of 480\;m from the interaction point (IP) and 5 m off the beam axis, consisted of a decay volume 35\;m long and with a surface area of 9\;m$^2$. We have modeled the detector following~\cite{Dobrich:2015jyk} and recast the analysis of~\cite{CHARM:1985anb}, in which an axion-like particle (ALP) decaying into a pair of photons was searched for. Since the analysis required only a single electromagnetic shower, it can be applied to the decay $N_2 \to N_1 \gamma$. We compute the number of events expected at CHARM following the equations that will be described in more details in Sec.~\ref{sec:SHiPFaser}, see Eq.~\eqref{eq:Nsignal} and subsequent ones. In our analysis, we simulate the production of $N_1 N_2$ pairs from the decay of the mesons $\rho$, $\omega,$ $J/\Psi$  and $\Upsilon$ using {\tt PYTHIA 8.3}~\cite{Sjostrand:2007gs,Bierlich:2022pfr},
finding the following production multiplicities: $N_{\rho}=0.58,$ $N_{\omega}=0.57,$ $N_{J/\Psi}=4.7\times10^{-6}$ and $N_{\Upsilon}=2.2\times10^{-9}$, see Sec.~\ref{sec:SHiPFaser} for their definition. 
Then, we require the energy of the photon in the laboratory frame to satisfy $E_\gamma \geq 1$ GeV and, following~\cite{CHARM:1985anb}, we take a signal acceptance of 51\%. The number of protons-on-target (POT) is taken to be $N_{\rm POT} = 2.4 \times 10^{18}$. 
Since no signal events were observed in the search of~\cite{CHARM:1985anb}, we set an upper limit at 95\% confidence level (CL) of  $N_{\rm signal} = 3$. The region excluded by the CHARM experiment is shown in Fig.~\ref{fig:sensitivities}\,,~\ref{fig:sensitivitiesEcut} and Fig.~\ref{fig:LHCexp}.

\paragraph{NuCal:}
In the $\nu$-calorimeter I experiment (NuCal), a 70\;GeV proton beam from the U70 accelerator was dumped on an iron target. The detector consisted of a cylindrical decay volume 26\;m long and with a diameter of 2.6\;m, placed at 64\;m from the iron target. We implement such geometry accepting $N_2$ events with a maximum angle of 0.014 rad from the beam axis. To set a limit on the parameter space of our scenario, we simulate $N_1 N_2$ production from $\rho$ and $\omega$ decays\footnote{We checked that production from heavier vector meson decays is negligible.} using {\tt PYTHIA 8.3} obtaining the following production multiplicities: $N_{\rho}=0.30,$ $N_{\omega}=0.30.$ Then, we follow the analysis in~\cite{Blumlein:2011mv}, requiring the photons produced in the $N_2 \to  N_1 \gamma$ decay to satisfy two conditions: their energy in the laboratory frame must be $E_\gamma \geq 3$ GeV, while their angle with respect to the beam axis must satisfy $\theta_\gamma < 0.05$ rad. After these selection cuts, 5 events were observed, with an estimated background of 3.5 events from the simulated neutrino interactions in the detector~\cite{Blumlein:1990ay}. Given these numbers, assuming Poisson likelihood we set a 95\% CL upper limit of $N_{\rm signal}\sim 7.1$~\cite{Blumlein:2011mv}. The region excluded by NuCal is shown in Fig. \ref{fig:sensitivities}\,,~\ref{fig:sensitivitiesEcut} and \ref{fig:LHCexp}.

\paragraph{NA64:}
In the NA64 experiment, an electron beam of 100\;GeV was dumped on a lead target. We have considered the analysis of~\cite{NA64:2020qwq}, in which an ALP decaying into a pair of photons was searched for. Since the two photons are too collimated to be distinguished, the final state was reconstructed as a single photon, allowing us to reinterpret  this search. In this case the $N_1 N_2$ pair is produced via photon bremsstrahlung. Using $2.84\times 10^{11}$ electrons-on-target~\cite{NA64:2020qwq}, we find that the number of events produced at NA64 is too small to put any bound on the parameter space of the model.

\subsection{Colliders}
We now analyze the bounds enforced by collider experiments, by considering searches performed at LEP, BaBar and LHC. In this case different searches apply, depending on whether the $N_2 \to N_1 \gamma$ decay is prompt, {\it i.e.} it happens at a distance smaller than $\sim 1\;$mm from the IP, displaced, {\it i.e.} it happens within $\sim 1\;$mm and $\sim 1\;$m, or else is detector-stable, {\emph{i.e}} when the decay happens at a distances greater than $\sim 1\;$m.
For a $2\to2$ scattering, in terms of $\hat{s} = (p_{N_1} + p_{N_2})^2$, we have
\be
\beta_{N_2} \gamma_{N_2} = \frac{\sqrt{\hat{s}}}{2 m_{N_2}} \sqrt{1 + \frac{(m_{N_2}^2 - m_{N_1}^2)^2}{\hat{s}^2} - \frac{2 \,(m_{N_1}^2 + m_{N_2}^2)}{\hat{s}}}, 
\ee
while the $N_2$ lifetime is given by $\tau_{N_2} = \Gamma(N_2 \to N_1\gamma)^{-1}$, with the decay width of Eq.~\eqref{eq:N2decay}. 

In the case of prompt decays, we consider two analyses: one from LEP~\cite{DELPHI:1996drf} and one from BaBar~\cite{BaBar:2017tiz}. In the LEP analysis, data were taken at various center of mass energies around the $Z$ peak. We employ the largest dataset, taken at $\sqrt{\hat s}=91.26\;$GeV with an integrated luminosity of 52.462 pb$^{-1}$. We have simulated our signal at the parton level by using {\tt MadGraph5\_aMCNLO}~\cite{Alwall:2014hca}. We enforce the analysis selections by requiring a single photon with $|\cos\theta_\gamma| < 0.7$ and considering two signal regions. In the first one a minimum energy of the photon was required $E_\gamma > 22$ GeV, and no events were observed. In the second region the cut is loosened to $E_\gamma > 3$ GeV, with 73 observed events and $72\pm5$ expected SM events. 
Therefore, we set a 95\% CL upper limit on the number of signal events $N_{\rm signal} =3$ for the first signal region and $N_{\rm signal} \sim 22$ for the second one.
For the weakly coupled normalization of the operator shown in Eq.~\eqref{eq:dipole_op}, the stronger bound corresponds to $\Lambda\lesssim17-50$\; GeV for $\delta = 0.1 - 1.$ However, these values of $\Lambda$ lie outside the range of validity of the EFT, implemented as in Eq.~\eqref{eq:validity_EFT}, therefore we conclude that no meaningful constraints on the cutoff scale $\Lambda$ can be set. 

We then move to the analysis of BaBar in Ref.~\cite{BaBar:2017tiz}, which derived bounds on single photon events produced in association with an invisibly decaying dark photon.
The selection of signal events makes use of a multivariate Boosted Decision Tree discriminant. Given the complexity of the analysis, we adopt a simplified strategy to estimate the constraint.
We use {\tt MadGraph5\_aMCNLO} to simulate a sample of $e^+ e^- \to N_1 N_2$ events at the center of mass energies corresponding to the $\Upsilon$(2s), $\Upsilon$(3s) and $\Upsilon$(4s) resonances, considering the luminosities reported in~\cite{BaBar:2017tiz}. We enforce the selections $-0.4 < \cos\theta_\gamma < 0.6$ and $E_\gamma > 3\;$GeV, corresponding to the LowM region of Tab. 1 of~\cite{BaBar:2017tiz}. To extract a bound, we focus on the 
loose ${\cal R}'_L$ selection of~\cite{BaBar:2017tiz}, and assume an equal number of observed and background events. This allows us to set an upper limit at 95\% CL of $N_{\rm signal} \sim 28$. The excluded region is largely independent of $m_{N_1}$ but depends quite strongly on $\delta$, since for $\delta \ll 1$ the energy of the photon in the laboratory frame is too small to pass the 3 GeV cut, see Eq. \eqref{eq:Egamma_LAB}. We find that, for $\delta = (0.5-1)$, the bound extends up to $\Lambda \sim 60$ GeV, while for $\delta < 0.5$ the bound disappears.
Given the approximate nature of our computation, we do not show these results explicitly in our figures.

Turning to searches for displaced events, we have considered analyses from ALEPH, ATLAS, CDFII and DELPHI~\cite{ALEPH:2002doz,ATLAS:2022bsa,CDF:2007sit,DELPHI:2003dlq}.  
Their reinterpretation is generally less straightforward than the ones for prompt signatures, 
due to the need of cutting on additional quantities as the photon time of flight and pointing variable. We decide to firstly apply a simplified strategy, by only imposing energy threshold and angular selection cuts on the final state photon. In this way the limit that we extract will be stronger than the one  obtained by a full implementation of the analysis. Using this strategy we find that the 
strongest bounds come from the DELPHI search \cite{DELPHI:2003dlq}, see also \cite{Ask:2002nuv}. In this case we enforce $E_\gamma>10\;$GeV and $|\eta_\gamma|<4.04$, which corresponds to the angular coverage between $2^\circ$ and $178^\circ$ reported in the analysis. We have once again simulated $e^+ e^- \to N_1 N_2$ and the subsequent $N_2 \to N_1 \gamma$ decay using  {\tt MadGraph5\_aMCNLO}, with center of mass
energies between 180 and 209\;GeV, and with the corresponding luminosities as reported in \cite{DELPHI:2003dlq}. Following this approach, we obtain 95\% CL limits which are in the $20 - 40$\;GeV ballpark, for $\delta = 0.1 - 1$. Given these results, we avoid implementing the full selection for the displaced analysis, since the obtained limits are already to be 
discarded because they lie beyond the validity of the EFT.

Finally, when $N_1$ and $N_2$ are both detector-stable, 
we consider searches of mono-$\gamma$ with missing energy, and the LEP limits on the invisible $Z$ width.
 It turns out that the strongest bound comes from the latter. By requiring $\Gamma(Z \to N_1 N_2) < 0.56$\;MeV~\cite{Chu:2018qrm} we obtain $\Lambda \gtrsim 9$\;GeV, which again lies beyond the validity of the EFT\footnote{Notice that in any UV completion of the dipole operator new states with masses around the EFT cut-off scale will be present, among which there will also be  states with non-vanishing electroweak charges.
Therefore, if their masses are small enough, additional bounds, that we are not discussing, could arise from the on-shell production of these particles.
 }.

We conclude this section by observing that some of the searches mentioned above could put meaningful constraints on the parameter space of strongly coupled UV completions of the EFT dipole operator. Implicit in our identification of the EFT cut-off scale with $\Lambda$ in Eq. \eqref{eq:dipole_op}, is the hypothesis that the dipole operator is generated perturbatively at one loop level by some heavy states. An alternative possibility could be for the dipole operator to be generated by some strong dynamics, similarly to what happens for the neutron in QCD. 
In this case, adopting the convenient parametrization of the dipole operator ${\cal O}^5_{NB}=1/\Lambda^{\prime}\,\bar N^c \sigma^{\mu\nu} N B_{\mu\nu},$  one expects the EFT cut-off scale to be of the order of $\Lambda^{\prime} =\Lambda\, (16\pi^2 )/g_Y.$
The bounds discussed above from the LEP search \cite{DELPHI:1996drf} valid for prompt decays are simply rescaled into $\Lambda^{\prime} \gtrsim (8 - 23)$\;TeV for $\delta = (0.1 - 1)$, essentially independent of $m_{N_1}$.
Analogously, the simplified approach adopted for the DELPHI search \cite{DELPHI:2003dlq} for displaced decays leads to the constraint $\Lambda^{\prime} \gtrsim (9 - 18)$\;TeV, again for $\delta = (0.1 - 1)$, essentially independent of $m_{N_1}$. Clearly in this case a more thorough reinterpretation of the analysis will be needed, with respect to the simplified approach previously described. 
Finally, the bound from the invisible $Z$ decay width valid for the detector-stable case would read 4\;TeV. 
Clearly, these constraints are probing a relevant part of the parameter space lying inside the regime of validity of the EFT, i.e. $\sqrt{\hat{s}}<\Lambda^{\prime}.$ 
It is important to notice that such bounds might be quite at odds with the range of $N_1$ and $N_2$ masses to which we are interested in. For example in a QCD-like strongly coupled scenario, we expect $N_1$ and $N_2$ to emerge as baryons, with masses of order $\Lambda^{\prime}$ and not much lighter as it would emerge from our analysis. 
A possibility of having a composite state much lighter than $\Lambda^\prime$ could be envisaged in a scenario 
where a light baryon arises in order to match the anomaly of an unbroken global chiral symmetry in the UV, along the lines of~\cite{tHooft:1979rat,Dimopoulos:1980hn}. We are not aware of any realistic model realizing such a framework. For this reason, in the remainder of the paper, we will consider only the weakly coupled scenario of Eq. \eqref{eq:dipole_op}.

\subsection{Bounds from astrophysics and cosmology}
\label{sec:cosmo}

In addition to the bounds presented above, limits from astrophysics and cosmology may be important for the scenario that we are considering. The constraint which is more relevant for us comes from BBN. Although in our simplified scenario the dipole operator ${\cal O}_{NB}^5$ completely governs the $N_2$ decays, the fate of $N_1$ is determined by its
mixing with the active sector. Particularly dangerous is the situation in which the $N_1$ decays could potentially spoil the predictions of the standard BBN model \cite{Boyarsky:2020dzc,Bondarenko:2021cpc,Sabti:2020yrt}. There are two natural ways to avoid this bound. Either $N_1$ is stable on cosmological scales, or $N_1$ decays with $\tau_{N_1} \lesssim 10^{-2}$ s. In the first case, $N_1$ would be a dark matter candidate\footnote{The case in which a fermionic dark matter candidate $\chi$ interacts via a dipole operator $\bar{\chi}\sigma^{\mu\nu} \chi F_{\mu\nu}$ has been studied in Refs. \cite{Chu:2018qrm,Chu:2020ysb}. Our case would correspond to an {\emph{inelastic dark matter}} scenario in which the dipole interactions are of the form $\bar{\chi}_2 \sigma^{\mu\nu} \chi_1 F_{\mu\nu}$. In the $\delta \ll 1$ limit, we expect the phenomenology generated by ${\cal O}_{NB}^5$ to be qualitatively similar to the one studied in \cite{Chu:2018qrm,Chu:2020ysb}. On the other hand, in the opposite limit $\delta \gtrsim 1$, the phenomenology can be quite different and will be studied elsewhere.}. In the second case, it has been shown in \cite{Boyarsky:2020dzc} that the combination of limits from BBN and terrestrial experiments exclude $m_{N_1} \lesssim (0.4 -0.5)$ GeV, for $N_1$ mixing dominantly with $\nu_e$ or $\nu_\mu$,
while lighter masses can be allowed for mixing dominantly with $\nu_\tau$.
Since in the first case an important region of parameter space that can be tested by the experiments we consider
would be excluded, we turn to the case of dominant mixing with $\nu_\tau$. Can such mixing be obtained in a way which is compatible with neutrino mass generation? As shown in \cite{Bondarenko:2021cpc}, in a scenario with only two RH neutrinos this is possible for $m_{N_1} \gtrsim 0.5\;(0.1)$~GeV for normal (inverted) hierarchy. The situation becomes less constrained considering three RH neutrinos, since in this case we have explicitly checked that a dominant $N_1 - \nu_\tau$ mixing can be obtained, in a way compatible with the generation of neutrino masses, for $m_{N_1} \gtrsim 0.1\;$GeV, which is the range we consider. This can be obtained also by assuming a mass hierarchy $m_{N_1} \sim m_{N_2} \ll m_{N_3}$, {\it i.e.} in a situation in which the phenomenology is driven by $N_1$ and $N_2$ as the one we are considering by using the simplified scenario of Eq.~\eqref{eq:dipole_op}.
In what follows, we will always implicitly suppose this to be the case.
For what concerns $N_2$, in Figs.~\ref{fig:sensitivities},~\ref{fig:sensitivitiesEcut},~\ref{fig:LHCexp} we show contours of constant $N_2$ lifetime, in order to highlight the region of the parameter space where $N_2$ decays fast enough to avoid BBN bounds.

Limits derived from supernov\ae~may also be important. 
Light particles produced in the interior of supernov\ae, which reach temperatures of several tens of MeV, can escape from the star, therefore cooling  the system. From this argument, masses up to a few hundred MeV can be constrained.
In the $\delta \ll 1$ limit, we expect our scenario to be qualitatively similar to the one studied in \cite{Chu:2018qrm}, where the Authors focus on a dipole operator constructed with a single new Dirac fermion playing the role of the dark matter. They obtain bounds up to masses of $\sim 0.1\;$GeV.
In the mass range $1-100$ MeV, these limits exclude $2\;{\rm TeV} \lesssim \Lambda \lesssim 50\;{\rm TeV}.$ 
The constraints disappear for larger $\Lambda$ because the production inside the supernov\ae~is suppressed, while for smaller $\Lambda$ efficient scattering processes can partially trap the particles inside the system.
For the case $\delta \gtrsim 1$ we expect the situation to be qualitatively different from the one above. For small enough values of $\Lambda$, $N_2$ decays quickly, thus leaving a dominant population of $N_1$ inside the supernov\ae. 
The only relevant $N_1$ scattering process is $N_1\;{\rm SM} \to N_2\;{\rm SM}$, which however might be not kinematically allowed for large $\delta$, possibly making the bound disappearing at small values of $\Lambda $. 
This scenario has been studied in the context of inelastic dark matter with a dark photon mediator in~\cite{Chang:2018rso}, albeit by using a simplified and conservative approach. Given the complexity of performing a detailed analysis of the supernov\ae~bounds, and the fact that we expect these limits to affect only a quite limited region of parameter space, we defer a detailed study of this problem for future work.

\section{Projected sensitivity of the SHiP and FASER 2 experiments}
\label{sec:SHiPFaser}

In this section we study the prospects for detection of long-lived RH neutrinos at the proposed future experiments SHiP~\cite{SHiP:2015vad} and FASER 2~\cite{Feng:2017uoz,Feng:2022inv}. These experimental facilities have the capability to probe long-lived particles in a variety of hidden sector models~\cite{Ahdida:2704147,Feng:2022inv}. SHiP is a fixed-target experiment, based on a high intensity 400 GeV proton beam dumped on a heavy target. Instead, FASER 2 is an LHC experiment, which aims at exploiting the proton collisions occurring at $\sqrt{s} = 14\,{\rm TeV}$ during the High-Luminosity LHC (HL-LHC) program.
In this kind of experiments, RH neutrinos can be copiously produced  by the decay of mesons generated by the proton collisions. The decay of long-lived $N_2$ particles can then show up in dedicated detectors located around the IP of these experiments.
The expected number of signal events can be computed as:
\be\label{eq:Nsignal}
N_{\rm signal} = N_{\rm prod} \left\langle \,\, f_{\rm dec}\, \epsilon_{\rm det} \,\, \right\rangle\ ,
\ee
where $N_{\rm prod}$ is the total number of $N_2$ produced, $f_{\rm dec}$ corresponds to the probability for $N_2$ to decay inside the detector volume, and $\epsilon_{\rm det}$ accounts both for the efficiency for the reconstruction of the events, that we simply take as 100\%, and selection cuts. Finally, $\langle \cdot \rangle$ indicates a statistical average, that we define in the following.
We consider the production of RH neutrinos from the decay of the following mesons: $M=\rho,\omega,J/\Psi, \Upsilon.$ The number of $N_2$ produced then reads: 
\be
\label{eq:NprodSHiP}
N_{\rm prod}= \sum_{M} N_{\rm POT}\,N_M\, {\rm BR}(M\rightarrow N_1 N_2),
\ee
where $N_{\rm POT}$ is the total number of collected protons on target, $N_{M}$ is the average number of mesons $M$ produced per proton interaction, and ${\rm{BR}}(M\rightarrow N_1 N_2)$ is the branching ratio of the decay of the meson $M$ into RH neutrinos, computed in App.~\ref{App:widths}. 
Similarly, the number of RH neutrinos produced by the decay of mesons at the LHC is given by:
\be
\label{eq:NprodLHC}
N_{\rm prod}^{\rm LHC} = \sum_{M} \sigma_{\rm ine}\,\mathcal{L}\,N_M\, {\rm BR}(M\rightarrow N_1N_2),
\ee
where $\mathcal{L}=3\,{\rm ab}^{-1}$ is the integrated luminosity at the HL-LHC, and $\sigma_{\rm ine}=79.5\,\rm{ mb}$ is the inelastic proton-proton cross-section~\cite{TOTEM:2017asr}.
The quantity $f_{\rm dec}$ is:
\be
\label{eq:fdec}
f_{\rm dec} =  e^{-L_{\rm entry}/L_{N_2}} - e^{-L_{\rm exit}/L_{N_2}},
\ee
where $L_{\rm entry}$ ($L_{\rm exit}$) is the distance between the IP where the $N_2$ particle is produced, and the point at which $N_2$ enters (exits) the detector.
Finally, $L_{N_2}$ is the decay length of $N_2$ in the laboratory frame, given by $L_{N_2}= \beta_{N_2}\gamma_{N_2} c \tau_{N_2}$.
Our calculations are based on simulations of the production of mesons performed with {\tt{PYTHIA 8.3}} and {\tt EPOS-LHC}~\cite{Pierog:2013ria}. More details will be given in the following. 
From these simulations we obtain a sample of mesons events, and we compute the associated multiplicities $N_{M}$. Then, for each meson event in the sample, we simulate its decay in $N_1N_2$ pairs. These data are used to statistically evaluate Eq.~\eqref{eq:Nsignal}, averaging ($\langle \cdot \rangle$) Eq.~\eqref{eq:fdec} over all the possible kinematical configurations of the $N_2$ particles in our sample.
Finally, we impose a minimum energy of the photon produced in the decay $N_2\rightarrow N_1+\gamma.$ We compute the efficiency of this selection cut, $\epsilon_{\rm det}$, using the $N_2$ events in our sample, simulating the $N_2$ decays, and selecting the events for which the photon energy in the laboratory frame is larger than a threshold $E_{\rm cut}.$
The procedure explained in this section has been used to compute the number of signal events also at the CHARM and NuCal experiments described in Sec.~\ref{sec:currentbounds}.

\subsection{SHiP}

The SHiP fixed-target experiment aims at accumulating $N_{\rm POT}=2\times10^{20}$ protons on a target composed by Molybdenum and Tungsten in 5 years of operation. A description of the experiment can be found in~\cite{SHiP:2021nfo}. The decay volume of the detector has a length of 50 m and it is located at $\sim$45 m from the proton target. A spectrometer and a particle identification system with a rectangular acceptance of $5\times 10\,{\rm m}^2$ are placed behind the decay volume. The rectangular face of the decay volume closer to the IP has a size of $1.5\times4.3\,{\rm m}^2$. Following these specifics, we approximate the detector as a cylinder with an opening angle of 31.8 mrad.

The production of the different mesons at SHiP is based on simulations of proton-proton collisions performed with {\tt{PYTHIA 8.3}}. For the $\rho$ and $\omega$ mesons we obtain the production multiplicities $N_{\rho}=0.58$ and $N_{\omega}=0.57$, in good agreement with previous results present in the literature~\cite{Dobrich:2019dxc,Darme:2020ral,Chu:2020ysb,Bertuzzo:2020rzo,SHiP:2020vbd}. 
We assume the same production rate for proton interactions in the target material of SHiP. In principle a dependence on nuclear target is expected, however detailed simulations or measurements are needed to fully capture these effects.
Instead for the $J/\Psi$, we normalize our simulation in order to reproduce the total number of mesons predicted by the SHiP collaboration in~\cite{SHiP:2018xqw}: $N_{J/\Psi}=2\times X_{\bar{c}c}\times f(q\rightarrow J/\Psi)\times f_{\rm cascade}.$ The $\bar{c}c$ production fraction is $X_{\bar{c}c}=1.7\times10^{-3}$, the $J/\Psi$ production fraction is $f(q\rightarrow J/\Psi)=0.01$ and the enhancement from cascade events is $f_{\rm cascade}=2.3$.
Finally, the production rate of the $\Upsilon$ mesons is directly obtained using {\tt{PYTHIA 8.3}}, as for the case of the $\rho$ and $\omega$, since detailed simulations of the production at SHiP are not available. We find\footnote{We use the same multiplicities of $\rho,$ $\omega$ and $\Upsilon$ for CHARM and SHiP since the energy of the proton beam is the same and we have neglected medium dependent effects.}  $N_{\Upsilon}=2.2\times10^{-9}$.

Currently, there are no studies of the background rates at SHiP for the single photon signature arising in our scenario. 
By assuming that the backgrounds can be reduced at a negligible level as it happens for other searches, see {\it e.g.}~\cite{SHiP:2018xqw}, the 95\% CL upper limit on the number of signal events is $N_{\rm signal}= 3.$
For a more conservative approach we follow~\cite{Magill:2018jla}, which estimated $\sim1000$ background events after rescaling 
the background events observed at the NOMAD detector by the number of POT in the two experiments.
This number will likely be reduced by vetos, as noticed in~\cite{Magill:2018jla}.
Assuming Poisson likelihood, we set a 95\% CL upper limit on the number of signal events of $N_{\rm signal}\sim 63.8$.
Finally, we impose a minimum energy of the photon $E_{\rm cut}=1\,{\rm GeV},$ which is a reasonable threshold for SHiP~\cite{SHiPprivate}.
Summarizing, in Fig.~\ref{fig:sensitivities} we show sensitivity contours for the two choices of signal events discussed here. This corresponds to a range between an optimistic and a conservative assumption of the background level.

\subsection{FASER 2}
\label{sec:FASER}

The FASER collaboration has proposed to build a suite of forward detectors to be placed within the LHC environment along the beam axis, nearby the ATLAS experiment~\cite{Feng:2022inv}.
These experiments are dedicated to study several interesting topics, as the properties of neutrinos, QCD in the forward regime and new physics beyond the SM, including dark sectors.
Several small size pilot detectors have already been constructed, which are FASER~\cite{Feng:2017uoz,FASER:2018eoc}, FASER$\nu$~\cite{FASER:2019dxq,FASER:2020gpr} and SND@LHC~\cite{SHiP:2020sos}. However, to fully exploit the potential of the HL-LHC, a dedicated facility to host larger detectors is under study. In particular, we focus on the proposed FASER 2 detector, which is dedicated to the study of long-lived particles. According to current design, it will be placed at 620\;m from the IP, and it will have a cylindrical shape, with a radius of 1 m and a length of 10 m~\cite{Feng:2022inv}.
  
To simulate the production of the $\rho$ and $\omega$ mesons at the LHC, we use {\tt EPOS-LHC}, which has been tuned to forward LHC data. For the $J/\Psi$ and $\Upsilon$ mesons we again use {\tt{PYTHIA 8.3}}, rescaling its rates to match the production cross-sections as measured by the LHCb experiment~\cite{LHCb:2018yzj,LHCb:2015foc}. In addition, following~\cite{Foroughi-Abari:2020qar}, we modify the production rate as a function of the transverse momenta $p_T$ with respect to the default setup of {\tt PYTHIA 8.3}. Employing this procedure, we obtain a good agreement with the measured $p_T$ distributions~\cite{LHCb:2018yzj,LHCb:2015foc}.  The resulting production multiplicities in one hemisphere are $N_{\rho}=2.3,$ $N_{\omega}=2.2,$ $N_{J/\Psi}=5.0\times10^{-4}$ and $N_{\Upsilon}=6.1\times10^{-6}$.
In addition to the production of RH neutrinos from the decay of mesons, we include Drell-Yan processes $q\bar{q}\rightarrow\gamma/Z\rightarrow N_1 N_2.$ More details are provided in App.~\ref{App:futureLHC}. This production mechanism is however in most cases subdominant with respect the one from mesons decay.

FASER 2 will be sensitive to photon signals and, at the same time, it will have the capability to strongly reduce the relevant backgrounds, see the discussion in~\cite{Jodlowski:2020vhr} where single photon signals have been studied in the context of a model of sterile neutrinos coupled to active neutrinos via a dipole operator\footnote{See also~\cite{Dreiner:2022swd} for an analogous signature in the context of R-parity violating SUSY.}. Since a thorough simulation of the relevant backgrounds has not been performed yet, we decide to follow the strategy of~\cite{Jodlowski:2020vhr}, presenting our results as isocontours of $N_{\rm signal}= 3$ and $N_{\rm signal}= 30$ events. A cut on the energy of the photon $E_{\rm cut}>100\,{\rm GeV}$ has been employed in our analysis, again inspired by~\cite{Jodlowski:2020vhr}. 

\subsection{Results}
\label{sec:results}

\begin{figure}
	\centering
	\includegraphics[scale=0.56]{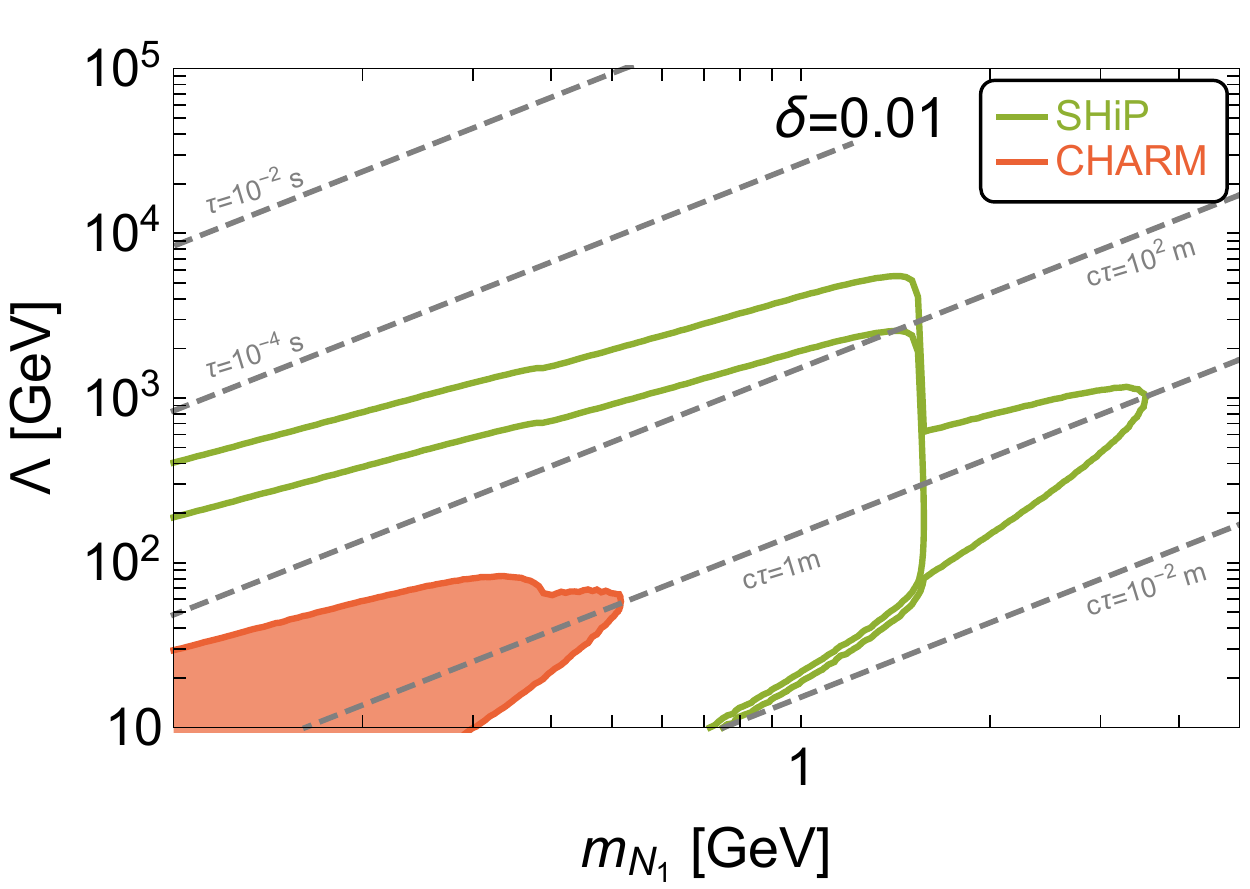}
	\quad
	\includegraphics[scale=0.56]{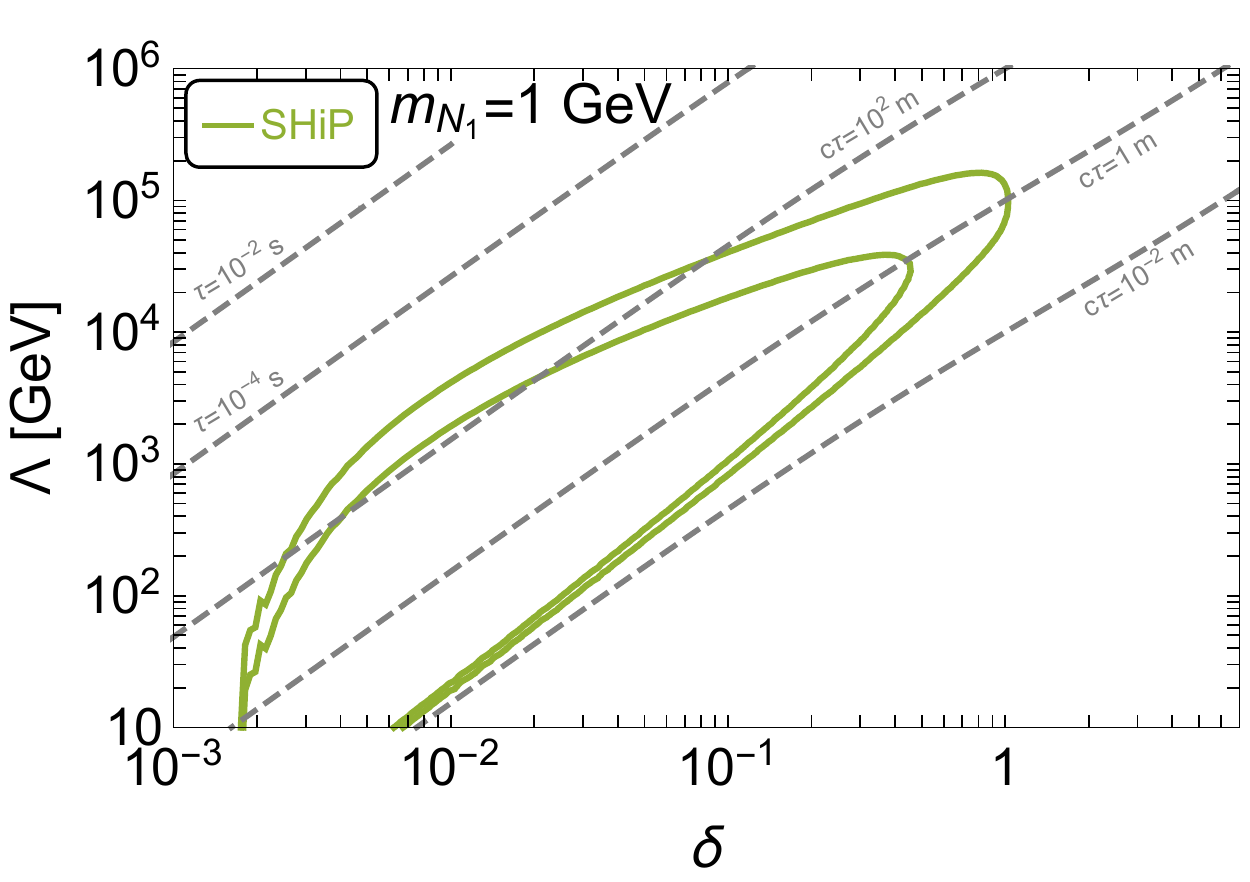}\\
	\quad
	\includegraphics[scale=0.56]{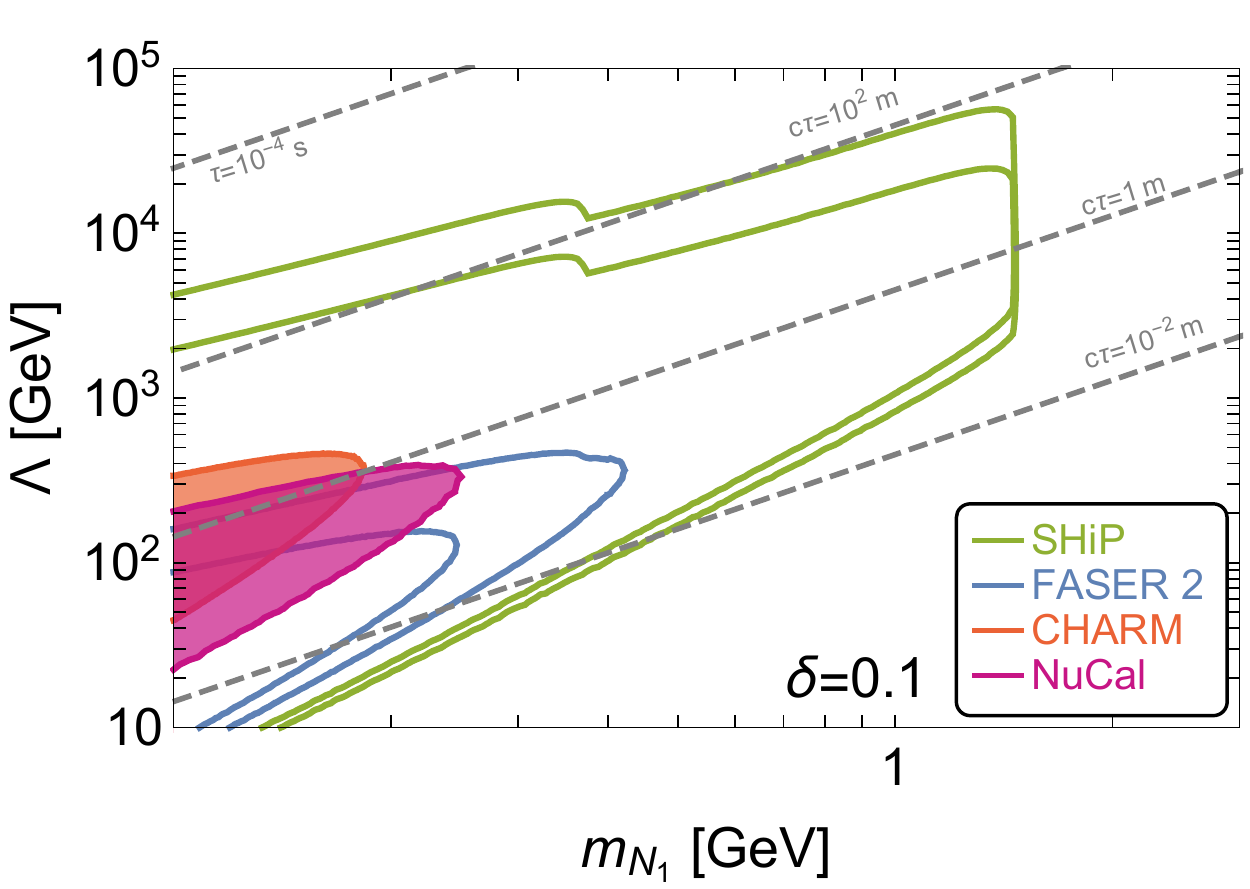}
	\quad
	\includegraphics[scale=0.56]{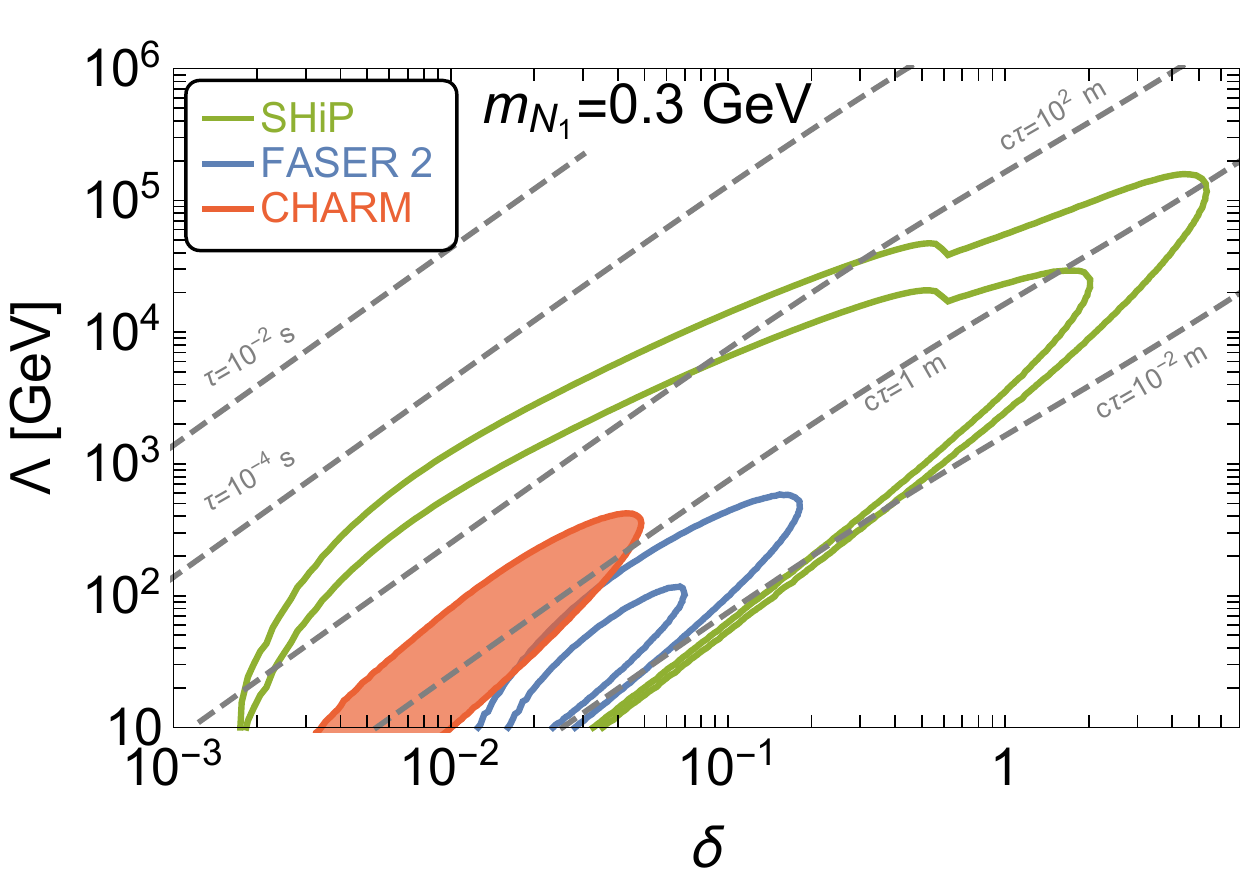}\\
	\quad
	\includegraphics[scale=0.56]{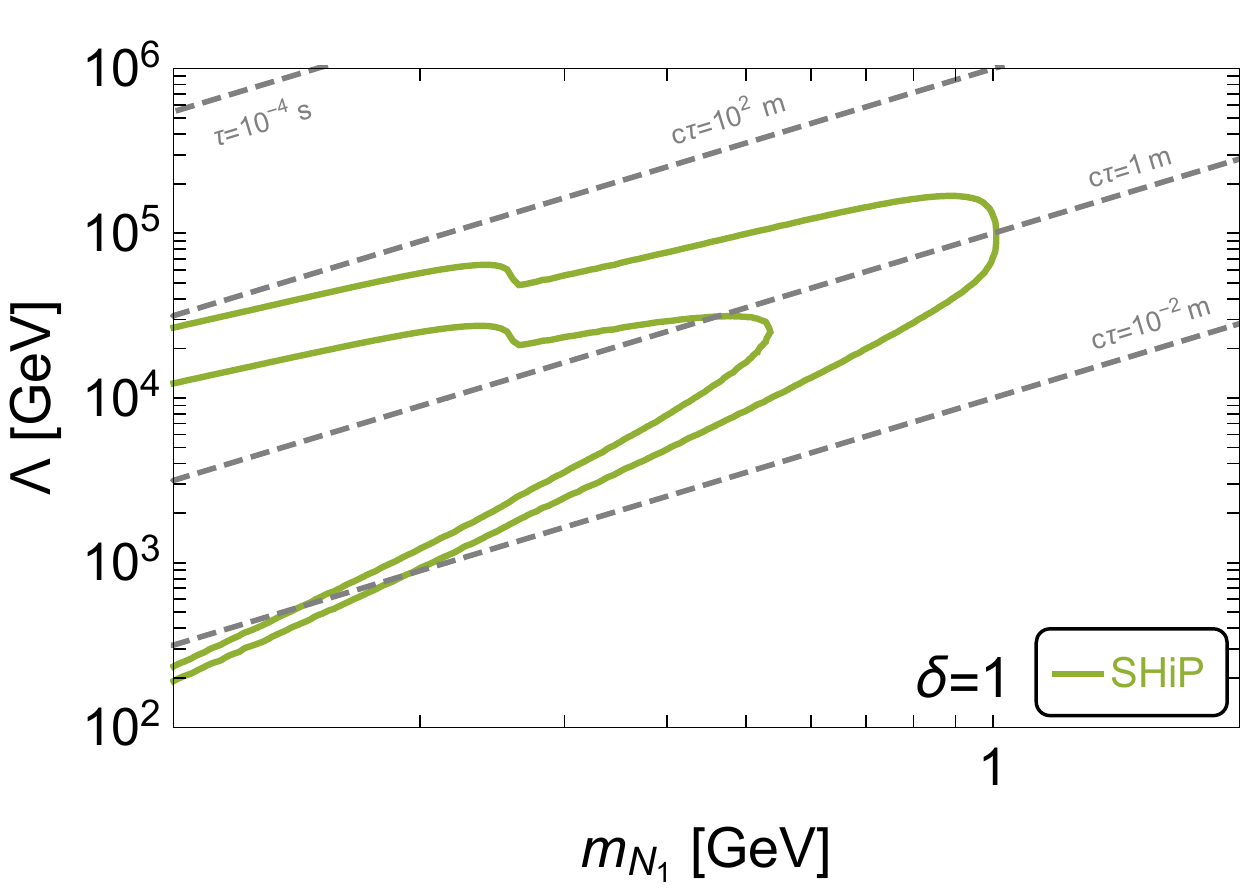}
	\quad
	\includegraphics[scale=0.56]{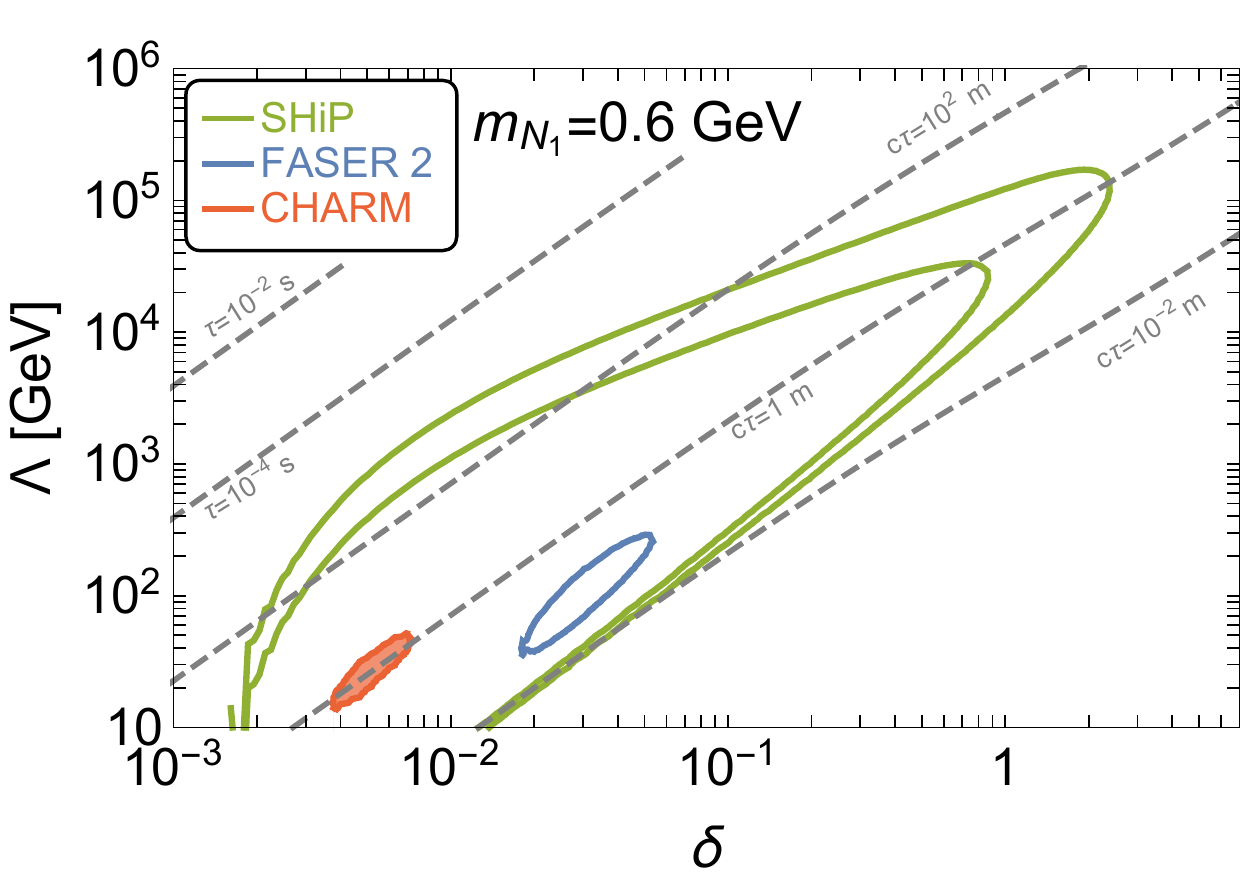}
	\caption{Green and blue lines are the sensitivity reach of the SHiP and FASER 2 experiments. For SHiP we show isocontours of $N_{\rm signal}=3$ and $N_{\rm signal}=63.8.$ For FASER 2 we show isocontours of $N_{\rm signal}=3$ and $N_{\rm signal}=63.8$, see Sec.~\ref{sec:SHiPFaser} for more details. The orange and magenta shaded regions are excluded by the CHARM and NuCal experiments. The dashed lines are contours of constant $N_2$ lifetime or proper decay length. We fix $\alpha=\pi/2$.}
	\label{fig:sensitivities}
\end{figure}

\begin{figure}
	\centering
	\includegraphics[scale=0.37]{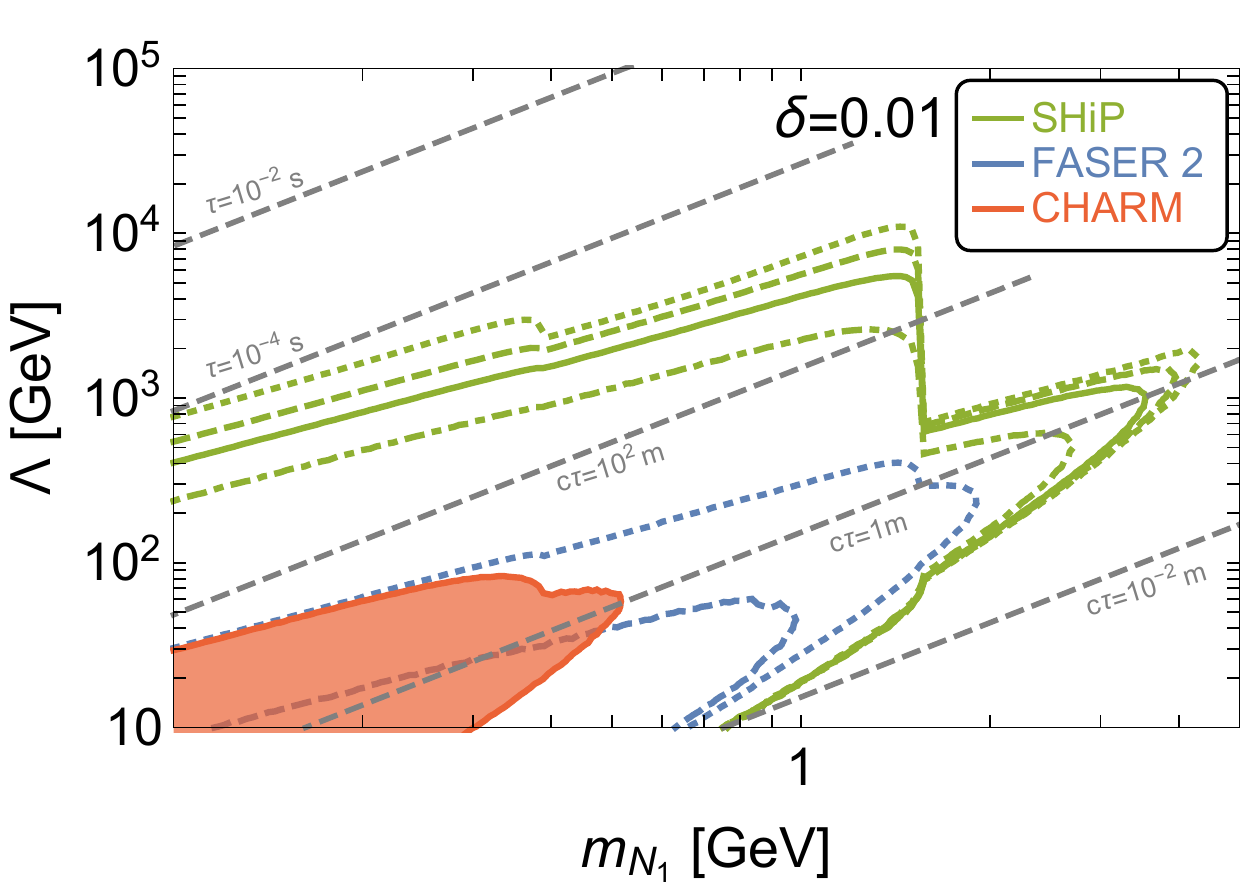}
	\quad
	\includegraphics[scale=0.37]{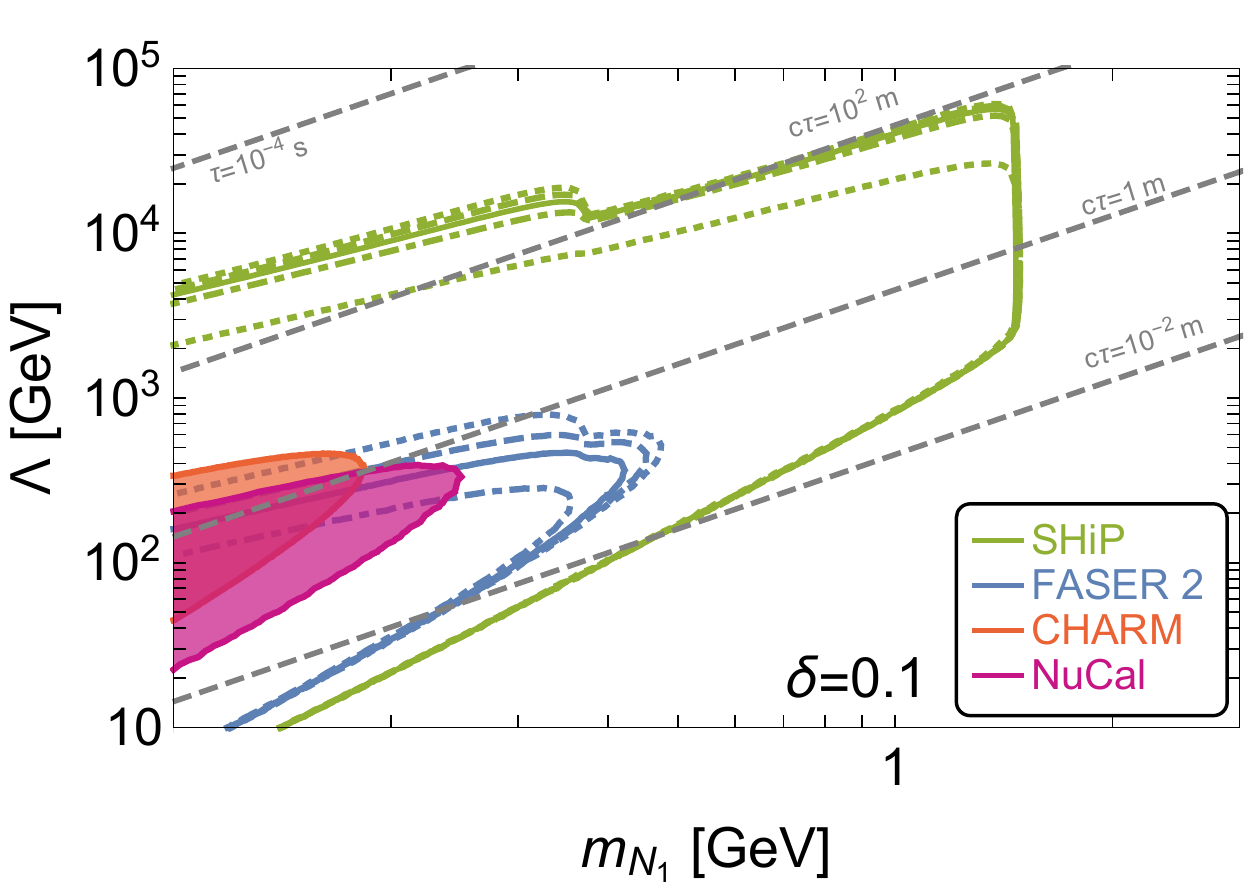}
	\quad
	\includegraphics[scale=0.37]{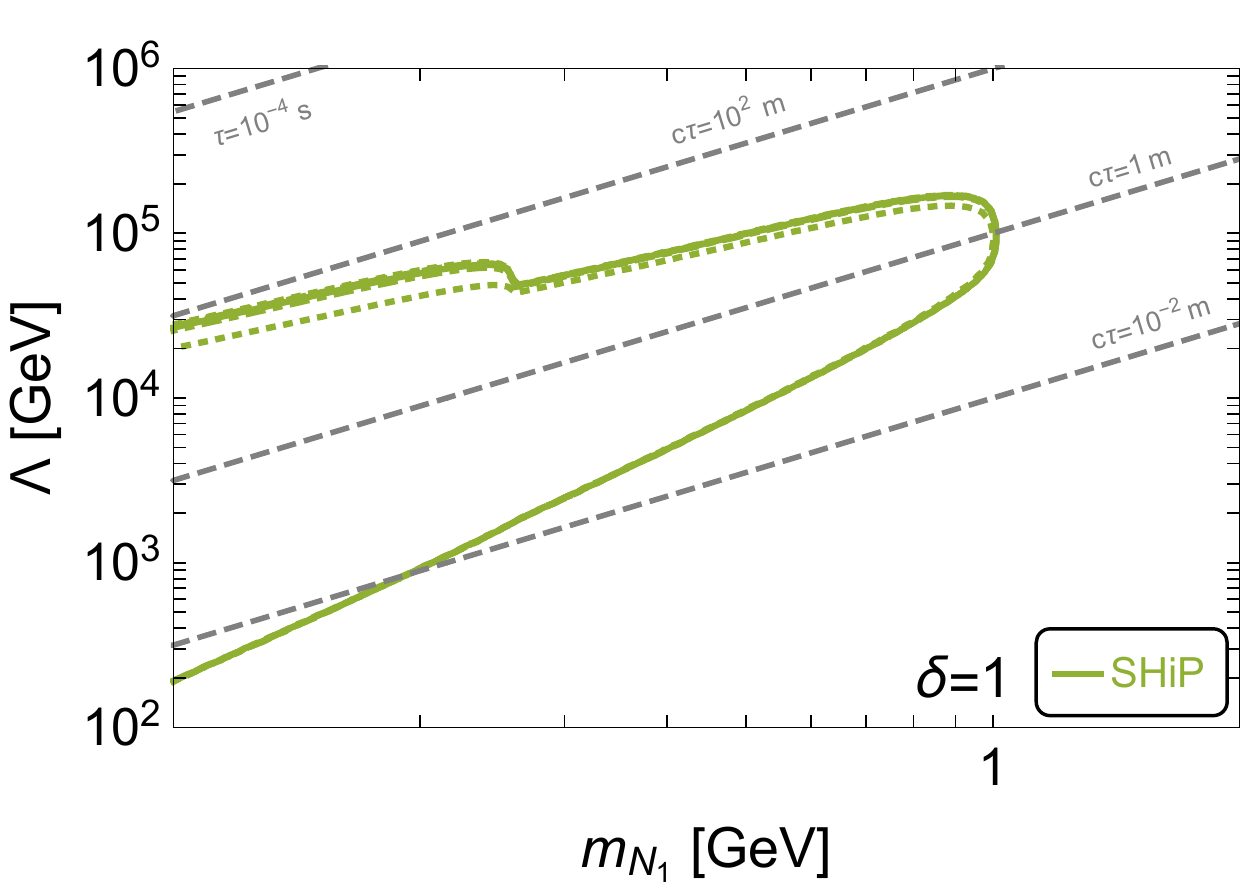}
	\caption{Isocontours of $N_{\rm signal}=3$ for SHiP (green lines) and FASER 2 (blue lines). For SHiP dotted, dashed, solid, dot-dashed and dotted lines are  for $E_{\rm cut}=0.1,0.5, 1, 2, 10$ GeV respectively while for FASER 2 dotted, dashed, solid and dot-dashed lines are for $E_{\rm cut}=10, 50, 100, 200$ GeV respectively.
	The CHARM and NuCal regions and the gray lines are as in Fig.~\ref{fig:sensitivities}.
We fix $\alpha=\pi/2$.}
	\label{fig:sensitivitiesEcut}
\end{figure}

Our main findings are presented in Fig.~\ref{fig:sensitivities}, where we fix the phase to $\alpha = \pi/2$ to maximize the production rate from meson decays, see App.~\ref{App:widths}. 
 The results remain qualitatively the same for other choices.
We show two different slices of the parameter space: either we fix the mass splitting $\delta$ and we explore the plane $m_{N_1}-\Lambda$ or we fix the mass of $N_1$ and we project the results on the $\delta-\Lambda$ plane.
 In both cases we consider three benchmark scenarios, namely $\delta=0.01,0.1,1$ in the first case and $m_{N_1}=0.3, 0.6, 1\;$GeV in the second case.
As explained in Sec.~\ref{sec:SHiPFaser}, the sensitivities of SHiP and FASER 2 are computed for two numbers of signal events, corresponding to different  choices of the background rate at these experiments. The strategy followed to compute the regions excluded by CHARM and NuCal is detailed in Sec.~\ref{sec:currentbounds}. When the line associated to a specific experiment is missing in our plots, this means that the corresponding experiment has not enough sensitivity to probe the parameter space.

As evident in Fig.~\ref{fig:sensitivities}, for a mass splitting $\delta=0.1,$ both SHiP and FASER 2 will be able to extend the current limits from CHARM and NuCal, and
probe an uncharted region of the parameter space.
In particular the sensitivity of SHiP reaches $N_1$ masses around the kinematical threshold for production from the decay of the $J/\Psi$ meson, {\emph{i.e.}} $m_{N_1}\sim1.5$ GeV.
A more modest sensitivity is obtained for FASER 2. It is worth recalling that despite small values of $\Lambda$ are formally not excluded in the EFT of Eq.~\eqref{eq:dipole_op}, weakly coupled UV completions with $\Lambda\lesssim100$ GeV are likely already ruled out from direct searches of additional EW charged states.
The constraints from BBN on $N_2$ decays are of the order $\tau_{N_2} = {\cal O}(10^{-2} - 1)\;{\rm s}$, see Sec.~\ref{sec:cosmo}. Looking at the isocontour of $\tau_{N_2}$ in  Fig.~\ref{fig:sensitivities}, one can notice that these bounds are not overlapping with the sensitivities of SHiP and FASER 2.
For a larger mass splitting, $\delta=1,$ the region probed by SHiP tends to shift to larger values of $\Lambda,$ while FASER 2 can not probe this slice of the parameter space. 
This behaviour can be understood by recalling that increasing $\delta$ tends to reduce the lifetime of $N_2$, see Eq.~\eqref{eq:N2decay}. This can be compensated by increasing $\Lambda$ at the price, however, of reducing the production rate of $N_1 N_2$ pairs. 
The correlation between $\delta$ and $\Lambda$ can be appreciated in the plots with $m_{N_1}$ fixed. The different experiments that we have studied are probing proper decay length $c\,\tau_{N_2}\sim10^{-2}-10^3$~m.

In the case of a smaller mass splitting, as in the case of $\delta = 0.01$, the threshold on the energy of the photon plays an important role.
Small  $\delta$ reduces the energy of the photon, see Eq.~\eqref{eq:Egamma_LAB}. This implies that at FASER 2 most of the events do not satisfy the cut $E_{\rm cut}>100\,{\rm GeV}$ and therefore no sensitivity is obtained.
To highlight the role of the energy threshold, in Fig.~\ref{fig:sensitivitiesEcut} we show the isocontours of $N_{\rm signal}=3$ for different values of $E_{\rm cut},$
namely $E_{\rm cut}=0.1, 0.5, 1, 2, 10$ GeV for SHiP and $E_{\rm cut}=10, 50, 100, 200$ GeV for FASER 2.
While for $\delta=1$ the sensitivities are almost unchanged, for smaller $\delta$ the energy threshold has a significant impact. In particular, for $E_{\rm cut}\sim 10$ GeV and provided that background can be kept negligible, FASER 2 will be able to test up to $\Lambda\sim400$ GeV for $\delta = 0.01,$ to be compared with a zero sensitivity scenario with $E_{\rm cut}\sim 100$ GeV, shown in Fig.~\ref{fig:sensitivities}.

In addition to FASER 2, other proposed LHC detectors targeting long-lived particles, as ANUBIS, CODEX-b, FACET and MAPP, might potentially be sensitive to the mono-photon signature. We estimate the number of signal events expected at these experiments in App.~\ref{App:futureLHC}.
In App.~\ref{sec:e-recoil} we discuss potential constraints arising from electron recoil searches in fixed-target experiments, finding much weaker sensitivities than those presented in this section. 

Finally, before concluding, we shall mention that the RH dipole operator might also be tested by the currently operating $e^{+}e^-$ collider experiment Belle II~\cite{Belle-II:2022cgf}, and by the future neutrino experiment DUNE~\cite{DUNE:2022aul}.
Dedicated analyses are in order to investigate their sensitivities.

\section{Conclusions}
\label{sec:conc}

In this work we have studied the phenomenological consequences of a dipole operator between RH neutrino fields. This is described by the $\nu$SMEFT $d=5$ operator $\bar N_2 \sigma^{\mu\nu}N_1 B_{\mu\nu}$ and triggers the decay $N_2 \to N_1 \gamma$, which is the subject of our study.
Motivated by the current experimental and theoretical interest, we have focused on RH neutrino masses in the GeV range and considered the regime in which $N_2$ is long-lived, with a proper decay length of ${\cal O}(10^{-2}-10^3\,{\rm m})$, while $N_1$ is considered to be stable on these length scales.

More in details, we have firstly considered the existing bounds on this scenario, arising from terrestrial experiments like CHARM, NuCal and colliders, as well as constraints from cosmological considerations, in particular in relation to the Big Bang Nucleosynthesis epoch.

We have subsequently investigated the sensitivity of the future proposed experiments FASER 2 and SHiP.
In these facilities the RH neutrinos are produced in $N_1 N_2$ pairs through the dipole operator, either 
via meson decay or via direct production.
Then, RH neutrinos give rise to single$-\gamma$ events through $N_2\to N_1 \gamma$ decays, which can be detected by these experiments in a background controlled environment.

Our main results are summarized in Fig.~\ref{fig:sensitivities} where we show that SHiP will be able to probe ample regions of the parameter space not yet excluded by current data, testing Wilson coefficients up to $\Lambda \sim 10^5$ GeV, while the sensitivity of FASER 2 is more limited.
Given the early design stage at which these experiments are, and the preliminary nature of the background estimates for the scenario under consideration, we have then studied how different cuts on the photon energy enforced at the analysis level affect the sensitivity reach.
Our results are shown in Fig.~ \ref{fig:sensitivitiesEcut}. We found that relaxing the cut on the photon energy has a limited impact on the sensitivities predicted for SHiP, while for FASER 2 ampler regions of parameter space can be reached, provided that the background can be maintained at a negligible level. Finally, {in Fig.~\ref{fig:LHCexp} in the appendix, we also present results on the number of signal events expected at other future LHC experiments, namely ANUBIS, CODEXb, FACET, MAPP.}

In conclusion, our work provides a first realistic estimate on the reach of experiments targeting long-lived particles on the lowest dimensional effective dipole operator that appears in the minimal see-saw extension of the Standard Model.

\medskip

\section*{Acknowledgements}
The Authors thank Michael Albrow, Oleg Brandt, David Curtin,  Jonathan L. Feng, Eric Van Herwijnen, Martin Hirsch, Gaia Lanfranchi, Vasiliki Mitsou, Thomas Ruf, and Sebastian Trojanowskifor useful discussions regarding the LHC experiments considered in this work. EB acknowledges financial support from FAPESP under contract 2019/04837-9.
M.T. acknowledges the research grant ``The Dark Universe: A Synergic Multimessenger Approach No. 2017X7X85'' funded by MIUR, and the project ``Theoretical Astroparticle Physics (TAsP)'' funded by Istituto Nazionale di Fisica Nucleare (INFN).
The work of CT was supported in part by MIUR under contract PRIN 2017L5W2PT.


\appendix
\section{Projected sensitivity of other future LHC experiments}
\label{App:futureLHC}

\begin{figure}
	\centering
	\includegraphics[scale=0.56]{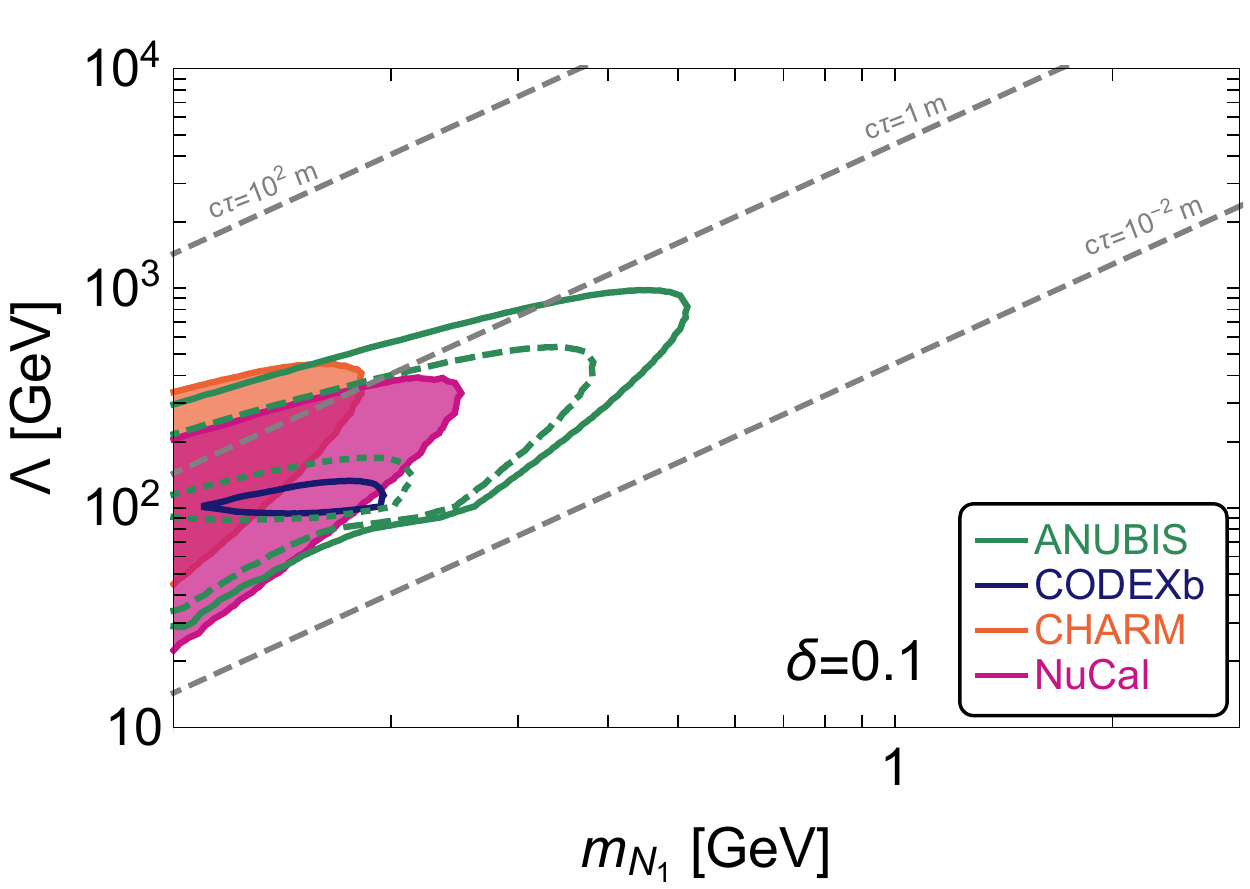}
	\quad
	\includegraphics[scale=0.56]{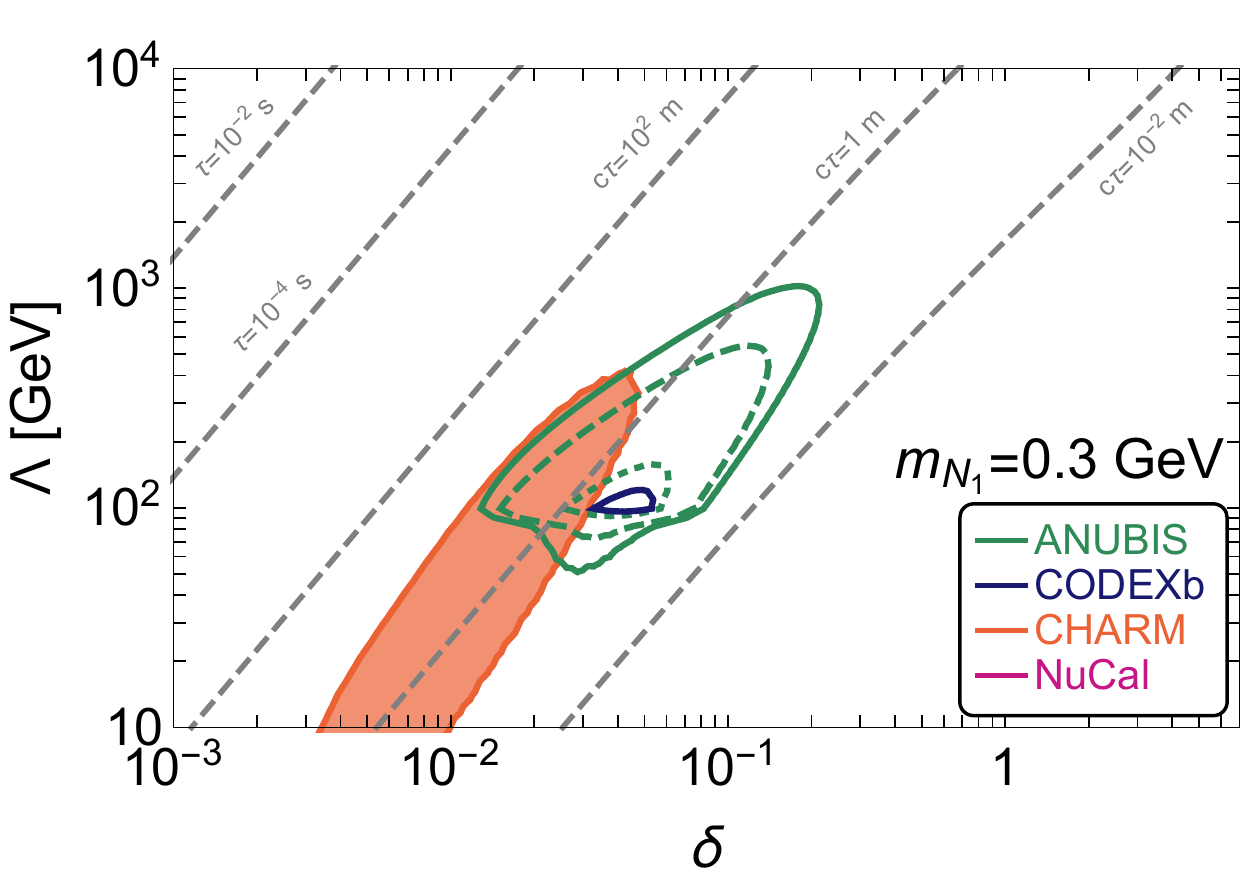}\\
	\quad
	\includegraphics[scale=0.56]{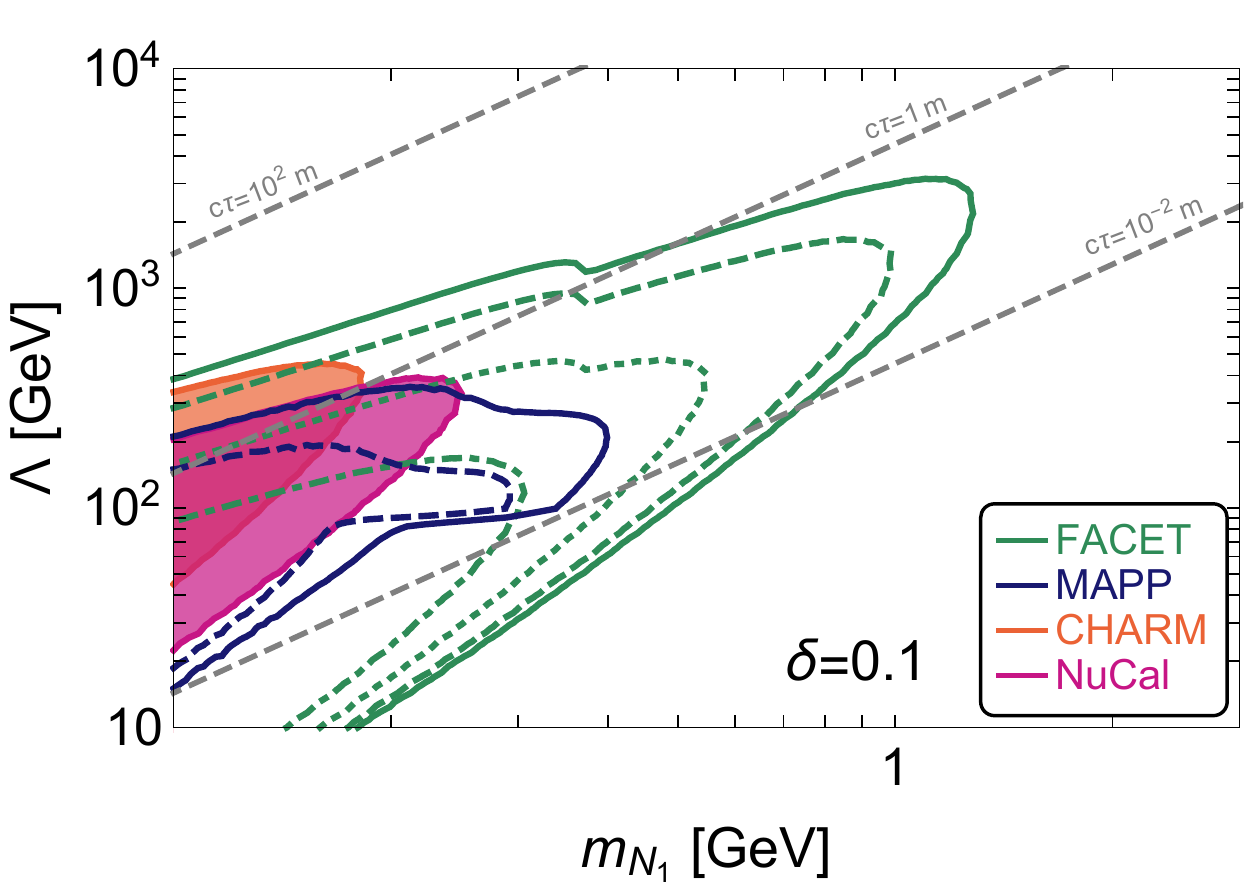}
	\quad
	\includegraphics[scale=0.56]{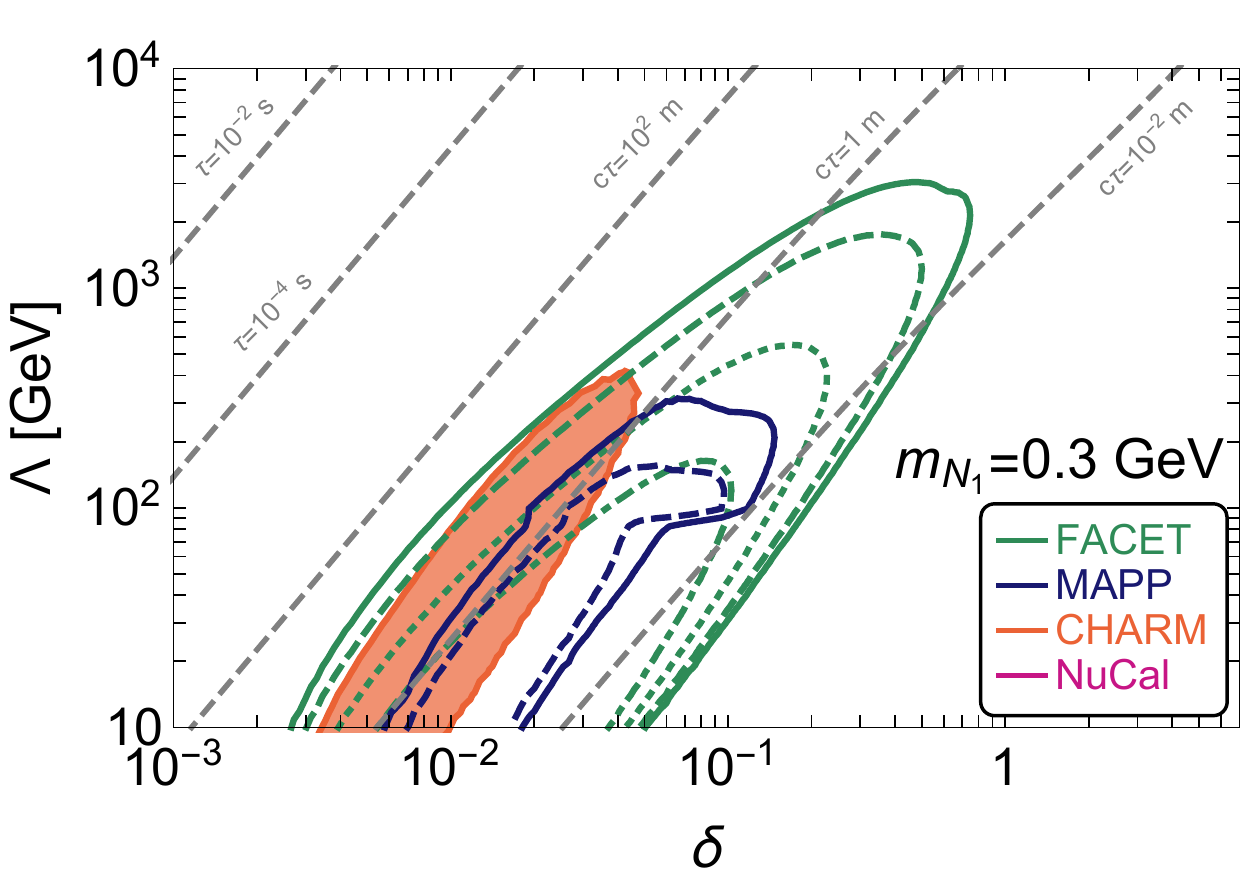}
	\caption{Sensitivity reach of the experiments ANUBIS, CODEX-b, CHARM and NuCal. Solid, dashed, dotted and dot-dashed lines correspond to $N_{\rm signal}=3,$ $N_{\rm signal}=10,$ $N_{\rm signal}=100$ and $N_{\rm signal}=1000.$
	The colored regions and the dashed lines are as in Fig.~\ref{fig:sensitivities}.
We fix $\alpha=\pi/2$.}
	\label{fig:LHCexp}
\end{figure}

In addition to FASER, several other LHC detectors dedicated to search for long-lived particles have been proposed in recent years: MATHUSLA~\cite{Chou:2016lxi,Curtin:2018mvb}, CODEX-b~\cite{Gligorov:2017nwh,Aielli:2019ivi,Aielli:2022awh}, AL3X~\cite{Gligorov:2018vkc}, MAPP~\cite{Staelens:2019gzt,Pinfold:2019zwp}, ANUBIS~\cite{Bauer:2019vqk} and FACET~\cite{Cerci:2021nlb}.
These facilities, to be placed around the LHC IP points, could potentially probe the radiative decay of the RH neutrinos that we are considering.
Concretely, we focus on CODEX-b, ANUBIS, MAPP and FACET, that, in principle, can have the potential to reconstruct photons (for CODEX-b this assumes an extension of the baseline design, according to~\cite{Aielli:2019ivi})\footnote{We thank members of the MATHUSLA, ANUBIS, MAPP and FACET collaborations for discussions on this point.}. For the single photon signature under scrutiny, the background rates at these experiments have not been computed, and a detailed discussion of their capability to reduce the relevant backgrounds is not currently available in the literature.
Given this limited information, the estimate of their sensitivity reach is quite uncertain. We show the region of the parameter space where the following numbers of signal events at these experiments are obtained: $N_{\rm signal}=3,10,100,1000.$ These results are intended to give an idea of potential sensitivity reach of these proposals, if the backgrounds are reduced to the appropriate rates.

For the calculation of the signal rate we follow the same procedure described in Sec.~\ref{sec:SHiPFaser}. The geometry of the different experiments and the cut on the photon energy $E_{\rm cut}$ are the same adopted in~\cite{Bertuzzo:2022ozu}.
We consider two mechanisms for the production of the RH neutrinos: meson decays, see Sec.~\ref{sec:FASER} for details, and Drell-Yan processes. For the latter, we employ {\tt MadGraph5\_aMCNLO} for our simulations. A couple of comments are in order.  As discussed in Sec.~\ref{sec:currentbounds}, we should require that the energy scale of this process is smaller than the cut-off of the EFT. Assuming a weakly coupled extension of the SM and couplings of $\mathcal{O}$(1), we impose $\sqrt{\hat{s}} < \Lambda$, where $\hat{s}$ is the Mandelstam variable associated to the process $q\bar{q}\rightarrow\gamma/Z\rightarrow N_1 N_2.$ In practice, from our {\tt MadGraph5\_aMCNLO} simulations, we select only the events satisfying this condition. In addition, we impose $\sqrt{\hat{s}} > 2\,{\rm GeV}$, in order to work in the regime of perturbative QCD.

The results are shown in Fig.~\ref{fig:LHCexp} for two different slices of the parameter space: fixing the mass splitting to $\delta=0.1$ or fixing the mass of the lightest RH neutrino to $m_{N_1}=0.3\;$GeV. For some experiments, FACET and ANUBIS, $N_{\rm signal}>10^2-10^3$ can be obtained in some parts of the parameter space, while more modest signal rates are obtained in other cases, as for CODEX-b and MAPP.

In general, we find that for the forward detector FACET and FASER 2, the production from meson decays is more relevant than the one from Drell-Yan processes. The opposite situation happens for the off-axis detectors ANUBIS and CODEX-b.
In some cases the sensitivity disappears at small $\Lambda$, see the bottom left panel of Fig.~\ref{fig:LHCexp} or the flattening of the curves in the other panels. This is due to the requirement $\sqrt{\hat{s}} < \Lambda$ that we impose in our simulation: for small enough $\Lambda$ most of the events in the simulation are rejected.

\section{Mesons decay into RH neutrinos}
\label{App:widths}

 For the decay $V \to N_1 N_2$ we need the following matrix element:
\be
\langle 0 | \bar{q} \gamma^\mu q | V(p) \rangle = f_V^q\, m_V\, \epsilon^\mu(p),
\ee
where $m_V$ is the vector meson $V$ mass, $\epsilon^\mu(p)$ its polarization vector and explicit expressions for the coefficients $f_V^q$ can be found in Appendix A of \cite{Bertuzzo:2020rzo}. Given the range of masses to which we are interested, in our computation we will consider only photon exchange. The explicit expression for the decay width is given by
\begin{align}
\begin{aligned}
\Gamma(V \to N_1 N_2) & = \frac{g_Y^2}{(16\pi^2 \Lambda)^2}\frac{(c_w\, Q_q\, e\, f_V^q)^2 m_V}{6\,\pi} \left(1-\frac{(m_{N_2} - m_{N_1})^2}{m_V^2} \right)^{1/2}\left(1-\frac{(m_{N_2} + m_{N_1})^2}{m_V^2} \right)^{1/2} \\
& \qquad \qquad \left(1 + \frac{m_{N_1}^2 + m_{N_2}^2- 6 m_{N_1} m_{N_2} \cos(2\alpha)}{m_V^2} -  2 \frac{(m_{N_2}^2 - m_{N_1}^2)^2}{m_V^4} \right) ,
\end{aligned}
\end{align}
where $c_w$ is the cosine of the weak angle, $Q_q$ the electric charge of quark $q$, in units of the electron's electric charge $e$. To produce our plots we set $\alpha = \pi/2$ to maximize the number of events, although the results remain qualitatively the same for other choices of the phase.

\section{Electron recoil searches}\label{sec:e-recoil}

The dipole operator of Eq.~\eqref{eq:dipole_op} induces an inelastic scattering processes between RH neutrinos and electrons, namely
\be
N_{2}\;e^{-}\to N_{1}\;e^{-}
\ee
and
\be
N_{1}e^{-}\to N_{2}e^{-} \ .
\ee
Note that the latter process is kinematically open only if the center of mass energy is sufficiently large, given that $m_{N_2}\ge m_{N_1}$. Such processes can give rise to a signal in experiments sensitive to $e$-recoils.
We have considered experimental searches at SHiP, CHARM II and DUNE and estimated present and future constraints.
These are fixed-target experiments, whose number of POT, detection efficiencies and cuts enforced in the analysis are recollected in Tab.~\ref{tab:e-recoil}.
\begin{table}[t!]
\begin{center}
\begin{tabular}{ccccc}
\toprule
Experiments & POT($10^{20}$) & $\epsilon_{{\rm eff}}$ & Cuts & References \\
\midrule
SHiP & 2 & $\sim1$ & $E_{R}\in[1,\,20]$ GeV,\,$\theta_{R}\in[10,\,20]$ mrad & \cite{SHiP:2015vad,Buonocore:2018xjk} \\
\midrule
CHARM II & 0.25 & $\sim1$ & $E_{R}\in[3,\,24]$ GeV,\,$E_{R}\theta_{R}^{2}\leq3$ MeV & \cite{CHARM-II:1989nic,CHARM-II:1994dzw} \\
\midrule
DUNE\,(10\,yr) & 11/yr & 0.5 & $E_{R}\in[0.6,\,15]$ GeV,\,$E_{R}\theta_{R}^{2}\leq1$ MeV & \cite{Brown:2018rcz,Hostert:2019iia} \\
\bottomrule
\end{tabular}
\end{center}
\caption{Summary of the main characteristics of the experiments that we considered. Here $E_{R}$ and $\theta_{R}$ are the recoil energy and the recoil angle with respect to the incoming neutrino's momentum of the scattered electrons, while $\epsilon_{{\rm eff}}$ is the detection efficiency of the signal.}
\label{tab:e-recoil}
\end{table}
The expected number of signal events is the sum of three contributions
\be
N_{{\rm sig}}=N_{12}+N_{21}+N_{212},
\ee
which are
\begin{itemize}
\item $N_{12}$: an $N_{1}$ particle produced by mesons decays produces an $e$-recoil signal in the detector through $N_{1}e^{-}\to N_{2}e^{-}$ scattering;
\item $N_{21}$: an $N_{2}$ particle produced by mesons decays produces an $e$-recoil signal in the detector through $N_{2}e^{-}\to N_{1}e^{-}$ scattering;
\item $N_{212}$: an $N_{1}$ particle produced by $N_{2}$ decay produces an $e$-recoil signal in the detector through $N_{1}e^{-}\to N_{2}e^{-}$ scattering.
\end{itemize}
Each term has been evaluated through a Montecarlo simulation of the process.
The $N_{1,2}$ neutrinos have been assumed to be produced from meson decays, and the fluxes of mesons have been simulated with {\tt{PYTHIA 8.3}}, as explained in the main text. The electron number density of the detectors has been obtained from the weight and the material of each experimental apparatus.

We have computed the differential cross section with respect to the recoil energy of the inelastic scattering processes, assuming initial electrons at rest. Explicitly
\begin{gather}
\frac{d\sigma}{dE_{R}}(N_{2}e^{-}\to N_{1}e^{-})=\frac{m_{e}}{2\pi}\left(\frac{e^{2}}{16\pi^{2}\,m_{e}\,P_{N_{2}}\Lambda}\right)^{2}f(m_{e}^{2}+m_{N_{2}}^{2}+2m_{e}E_{N_{2}},-2m_{e}E_{R}),\\
\frac{d\sigma}{dE_{R}}(N_{1}e^{-}\to N_{2}e^{-})=\frac{m_{e}}{2\pi}\left(\frac{e^{2}}{16\pi^{2}\,m_{e}\,P_{N_{1}}\Lambda}\right)^{2}f(m_{e}^{2}+m_{N_{1}}^{2}+2m_{e}E_{N_{1}},-2m_{e}E_{R}),
\end{gather}
where $P_{N_{1,2}}$ are the moduli of the RH neutrino spacial momenta, $E_{N_{1,2}}$ their energy, the recoil energy $E_{R}$ is the kinetic energy of the final electron and
\be
\begin{split}
f(s,t)=\frac{1}{t^{2}}\big\{&4(m_{N_{1}}m_{N_{2}}m_{e})^{2}-2m_{e}^{4}t+(t^{2}+2st)(m_{N_{1}}^{2}+m_{N_{2}}^{2}+2m_{e}^{2})\\
&-(t+2m_{e}^{2})(m_{N_{1}}^{4}+m_{N_{2}}^{4})-2st(s+t)-2m_{N_{1}}m_{N_{2}}t(t+2m_{e}^{2})\cos(2\alpha)\big\}
\end{split}
\ee
with $\lambda(x,y,z)=x^2 + y^2 + z^2 -2xy - 2yz - 2zx$.

We find the number of recoil events to be negligible for values of $\Lambda$ above $1\;$GeV, independently of the values of $m_{N_1}$ and $\delta$. We then conclude that searches through electron recoils do not impose relevant constraint on the $\nu$SMEFT parameter space.
 In the limiting case $\delta = 0$, we reproduce existing results available in the literature, which have focused on the case of elastic scattering processes~\cite{Chu:2020ysb}. 
\vspace{1cm}


\bibliographystyle{JHEP}
{\footnotesize
\bibliography{biblio}}
\end{document}